\documentclass[%
 reprint,
 amsmath,amssymb,
 aps,
prb,
]{revtex4-2}

\usepackage{graphicx}
\usepackage{dcolumn}
\usepackage{bm}
\usepackage{xcolor}
\usepackage{physics}
\usepackage[caption=false]{subfig}
\usepackage{ragged2e}
\usepackage{mwe}
\usepackage{tcolorbox}
\usepackage{bbold}
\usepackage{tkz-euclide}
\usepackage{wasysym}
\usepackage[T1]{fontenc}
\usepackage{blindtext, rotating}
\DeclareCaptionJustification{justified}{\justifying}
\renewcommand{\vec}[1]{{\boldsymbol #1}}

\newcommand{\nn}[1]{\langle #1 \rangle}
\newcommand{\rucl}[1]{$\alpha$-RuCl$_3$}
\newcommand{\uijop}[1]{\hat{u}^\alpha_{ij}}
\newcommand{\uij}[1]{u^\alpha_{ij}}

\definecolor{DarkGreen}{RGB}{1,50,32}
\begin{document}

\preprint{APS/123-QED}

\title{Raman spectroscopy of anyons in generic Kitaev spin liquids}

\author{Aprem P. Joy}
\author{Achim Rosch}%
\affiliation{%
	Institute for Theoretical Physics, University of Cologne, Cologne, Germany
}%
\date{\today}

\begin{abstract}
Optical probes have emerged as versatile tools for detecting exotic fractionalized phases in quantum materials. We calculate the low-energy Raman response arising from mobile, interacting Ising anyons (or visons)  in the chiral Kitaev spin liquid perturbed by symmetry allowed interactions - a phase relevant to \rucl. under a magnetic field. At zero temperature, the two-anyon continuum response shows a leading power-law scaling of the intensity near the onset of the signal: $I(\omega) \sim (\omega-E^0_{2\sigma})^{\frac{1}{8}}$ for linear and parallel-circular polarization channels, where  $E^0_{2\sigma}$ is the two-particle gap. Strong corrections due to short-range interactions arise at order $\frac{1}{4}$. For cross-circularly polarized channels, the scaling is given by $I(\omega) \sim (\omega-E^0_{2\sigma})^{|l\pm 1/8|}$, where the value of $l=0,1,2$ is determined by the number of minima in the single anyon dispersion. The exponents are directly related to the topological spin of Ising anyons $\theta_\sigma =\frac{\pi}{8}$, describing their exchange statistics. Our theory generalizes to spectral probes of anyonic quasiparticles with multiple band minima in other
quantum liquids. Interaction between anyons may also induce bound-states, resulting in sharp peaks that show strong polarization dependence.
\end{abstract}

\maketitle


\section{\label{sec:intro}INTRODUCTION}
Fractionalization of fundamental quantum numbers in many-body systems is one of the most remarkable discoveries of modern condensed matter physics. In such {\em topological} phases of matter, the elementary excitations may carry a fraction of the underlying electronic quantum numbers. In two-dimensions, as a consequence of long-range entanglement, such quasiparticles can also exhibit anyonic exchange statistics: they are neither fermions  nor bosons \cite{leinaas1977theory}. 

A well-known example is the fractional quantum Hall effect where the emergent quasiparticles not only carry a fractional electric charge but also are anyons with fractional statistics. Anyonic excitations also emerge in magnetic systems that realize the so-called quantum spin liquid states. A paradigmatic theoretical model is the Kitaev honeycomb model \cite{Kitaev06} which realizes a quantum spin liquid where the excitations are described by Majorana fermions coupled to a static $\mathbb Z_2$ gauge field. In the presence of a weak external magnetic field, the $\mathbb Z_2$ vortices, termed $visons$, become non-Abelian anyons.

The Kitaev spin liquid is at the forefront of the quest for QSLs and anyons. Besides its exact solvability, the model is also believed to be an approximate description of certain spin-orbit coupled quantum magnets \cite{jackeli,gammaRau,chaloupka2010kitaev}. The most notable of these being \rucl. and Na$_2$IrO$_2$ \cite{trebstreview,janvsa2018observation,nairo3}.  Although almost all the candidate materials order magnetically at low enough temperatures and thus fail to realize the Kitaev model, an intriguing scenario arises when the magnetic order is destroyed by an external magnetic field or higher temperatures.

Neutron scattering experiments done at temperatures above magnetic ordering have reported a highly disordered and broad response, which was attributed to the presence of fractionalized excitations \cite{proximate,fieldinduced}. Alternate explanations involving disorder or interactions, however, challenge these interpretations. The experimental observation of an approximately half-quantized thermal Hall effect (THE) in \rucl.  when the magnetic order is destroyed by an external magnetic field is a remarkable result that catapulted \rucl. into the spotlight \cite{kasahara1,kasahara2}. Such a half-quantized THE is a smoking gun signature of Majorana fermions and, if confirmed, would constitute a strong evidence for the realization of a Kitaev spin liquid. 
These developments, although encouraging, calls for a better understanding of Kitaev materials in order to make concrete experimental predictions.

Conventional experimental probes like neutron-scattering has been proposed to observe signatures of fractional quasiparticles in a variety of systems. An instructive work by Morampudi \textit{et al.} \cite{morampudi} argued that long-range statistical interaction between anyons must result in a characteristic power-law onset in spectroscopic response. 
Inelastic light scattering, in this context, offers a complementary tool to study quasiparticle excitations of a magnetic system, ranging from conventional magnons to fractionalized spinons \cite{wulferding2019raman}. In addition, the polarization degree of freedom (of light) allows one to directly probe the symmetry properties and chirality of excitations. Indeed, polarization resolved Raman scattering on quasi-1D spin chain materials have been successful in observing fractionalized spinon excitations, as well as various bound states of spinons \cite{choi2021bosonic,PhysRevB.53.R14733,gnezdilov2012phononic}. 

In two-dimensional systems, Raman scattering has been discussed as an effective probe to detect spinons and emergent gauge fields in $U(1)$ Dirac spin liquids \cite{raman_u1} and chiral spin liquids \cite{PhysRevB.106.064428}. 

Previous theoretical studies of Raman response on the pure Kitaev model predicted a (largely) polarization independent, gapless continuum response arising from the Majorana fermions alone \cite{PhysRevLett.113.187201}. This was later invoked to explain the observed temperature dependence of the Raman response of \rucl., which suggested the existence of fermionic quasiparticles \cite{sandilands2015scattering,nasu2016fermionic}. Raman experiments under a magnetic field, in contrast, have observed not only a gapped continuum response, but also multiple sharp peaks. Although the interpretation remains unclear, it was suggested to arise from the anyonic excitations of a chiral Kitaev spin liquid and their bound states \cite{wulferding2020magnon}. Recent experiments \cite{anuja1,anuja_chiral} on \rucl. conducted over a wide range of magnetic field strength have observed sharp peaks that show strong sensitivity to the handedness of light, suggesting the presence of chiral excitations in the system.

Given the wealth of experimental results, it is necessary to understand the direct signatures of the emergent excitations of the Kitaev liquid, if realized in a candidate material. 
A major hurdle to this is the inevitable presence of additional spin interactions that spoil the integrability of the model and introduce new dynamical degrees of freedom. In a previous publication \cite{prxvison}, we have analyzed, using a controlled perturbation approach, how anyonic excitations acquire dynamics in a generic Kitaev liquid, which are otherwise static in the pure model.

In this paper, we consider a generic Kitaev liquid with mobile anyons as the low-energy degrees of freedom, and study its Raman response. Microscopically, the interaction between electromagnetic fields and the emergent gauge field enters through spin-photon coupling \cite{Loudon_Fleury}. While in the pure Kitaev model, Raman operators only couple to the Majorana fermions, generic symmetry allowed Raman operators in a perturbed Kitaev model can excite the anyons in pairs, allowing for a direct probe of their dynamics, interaction, and statistics. A microscopic derivation of symmetry allowed Raman operator, beyond the conventional Loudon-Fleury theory was carried out recently by Yang \textit{et. al} in Ref.~\cite{nonLF_perkins}.

The paper is organized as follows. In Sec. \ref{sec:model}, we will briefly review the pure Kitaev model and its emergent anyonic excitations. In Sec \ref{sec:GKSL}, we discuss the low-energy dynamics of anyons in a generic Kitaev chiral spin liquid (gKCSL). The coupling between Raman operators and the anyonic excitations is discussed in Sec. \ref{sec:ramanop}. We then derive an effective Hamiltonian within the two-anyon subspace for a gKCSL in Sec. \ref{sec:effective}, which is used to obtain the Raman response to both linear and circularly polarized light presented in Sec. \ref{sec:spectrum}. Finally, we discuss the results and their implications for experiments on Kitaev materials.
\section{Model}
\label{sec:model}
\begin{figure}
	\centering
	\includegraphics[width=0.4\textwidth]{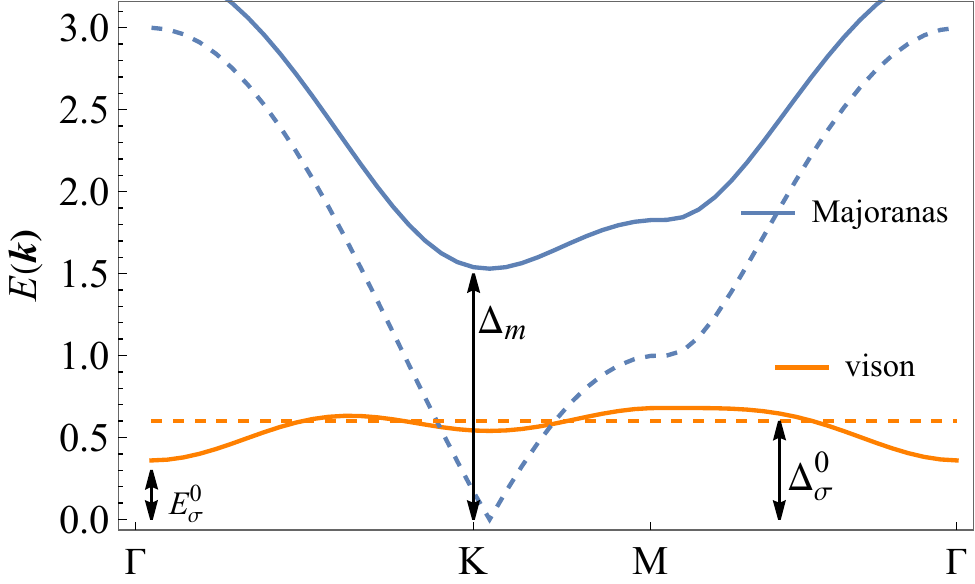}
	\caption{\textbf{Excitations in a generic Kitaev spin liquid in the chiral phase.} In the pure Kitaev model, elementary excitations are gapless Majorana fermions (dashed blue lines) and $\mathbb{Z}_2$ fluxes (visons) that are gapped and immobile (dashed orange lines). In a generic Kitaev liquid in a weak magnetic field, a topological gap $\Delta_m$ appears in the Majorana spectrum while the visons transform into Ising anyons that become dynamical degrees of freedom (solid lines).}
	\label{fig:bandsdemo}
\end{figure}
The Kitaev model is described by the following Hamiltonian defined on a honeycomb lattice of spin-half moments
\begin{align}
	H_K = K \sum_{\nn{ij}_\alpha}\sigma^\alpha_i \sigma^\alpha_j,
\end{align}
where $\alpha = x,y,z$ defines three types of links on the lattice, with vectors $\vec d_{\nn{ij}}$, as shown in Fig.~\ref{fig:geometry}. We fix $i \in $ sublattice A and $j \in$ sublattice B.
The exact solution,  à la Kitaev, can be obtained by mapping each spin operator to four Majorana fermions
\begin{align}
	\sigma^\alpha = i b^\alpha_i c_i.
 \end{align}
Identifying the conserved \textit{link} operator $\hat{u}_{\nn{ij}_\alpha} = i b_i^\alpha b_j^\alpha$,
the Hamiltonian transforms into a model of freely hopping Majorana fermions coupled to a $\mathbb Z_2$ gauge theory
\begin{align}
	H_K = K \sum_{\nn{ij}} i \uij{} c_i c_j.
\end{align}
Here, $\uij{}=\pm 1$ denote the eigenvalues of $\uijop{} $, forming the $\mathbb Z_2$ gauge variables.
The physical degrees of freedom of the gauge field, however, are the $\mathbb Z_2$ flux operators on each hexagonal plaquette, defined as
\begin{align}
	\hat W_p = \prod_{\hexagon} \sigma^\alpha \sigma^\alpha = \prod_{\hexagon} \uij{},
\end{align}
with eigenvalues $w_p=\pm 1$.  An elementary gauge field excitation thus corresponds to a single plaquette operator with eigenvalue $-1$. This defines a localized quasiparticle dubbed \textit{vison}, which carries the gauge flux. A vison acts as a source of $\pi$ flux for the `$c$' fermions ({\em matter} fermions). While local operators (e.g. $\sigma^x_i$) can only create visons in pairs, an isolated vison is a well defined quasiparticle with a finite energy cost $\Delta_v^0 \approx 0.15 |K|$, making them deconfined excitations. Due to the conservation laws $[H_K, \hat{W}_p]=0$ and $[H_K, \uijop{}]=0$, the emergent gauge field has no dynamics in the Kitaev model. This crucial property enables one to obtain the exact eigenvalues and eigenstates of the model by diagonalizing the matter Majorana Hamiltonian for each flux configuration after gauge fixing.

It can be shown that the ground state of the system lies in the sector with no visons and one can choose a gauge configuration with all $\uij{}=+1$ \cite{Kitaev06}. This in turn leads to a gapless Dirac spectrum for the matter Majoranas which describes the low-energy dynamics of the Kitaev model.
\subsection{Chiral spin liquid and Ising anyons}
In the presence of an external magnetic field, time reversal symmetry breaking induces a topological gap for the matter Majoranas, driving the system into a chiral spin liquid (CSL) phase \cite{Kitaev06}. As pointed out by Kitaev, projecting the effect of the magnetic field into the vison-free ground state of the mode, a mass term for the Majoranas arises at third order perturbation theory in the field. Remarkably, this preserves the exact solvability fo the model. Thus the low-energy Hamiltonian for a chiral Kitaev spin liquid takes the form
\begin{align}
	H_{CSL} = & H_K+ \kappa\sum_{[ijk]_{\alpha \beta \gamma}} \sigma^x_i\sigma^y_j\sigma^z_k,\\= &H_K +\kappa \sum_{[ijk]_{\alpha \beta \gamma}} i u^\alpha_{ij}u^\gamma_{{kj}} c_i c_k.
\end{align}
where $[ijk] = \nn{ij}_\alpha\nn{jk}_\beta$ denotes two adjacent links $ij$ and $jk$. We used the fermionic representation to go from the first to the second line of the above equation. The induced Majorana gap $\Delta_m \approx 6\sqrt{3} \kappa $ for small $\kappa/|K|$. 

 \begin{figure}
	\centering
	\begin{tikzpicture}
		\node[anchor=center] (image) at (0,0) {\includegraphics[width=0.45\textwidth]{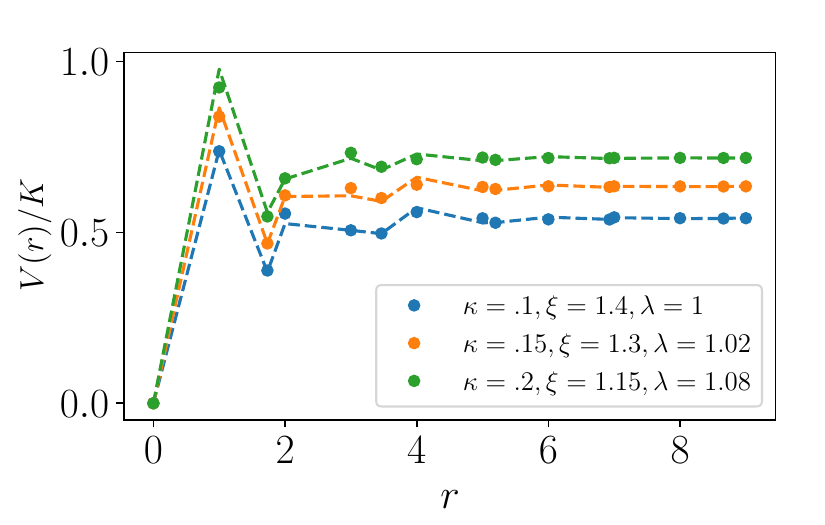}};
		\node[anchor=center] (text) at (-4,2.5) {\small{(a)}};
	\end{tikzpicture}

	\caption{Energy of a static (bosonic) Ising anyon pair as a function of their separation $r$, plotted for different values of $\kappa$. The dashed lines are obtained by numerically fitting the data to the functional form in Eq.~\eqref{eq:interaction} with fitting parameters $\xi$ and $\lambda$ shown in the figure legends. The values are obtained for a system of linear size $L=34$ with periodic boundary conditions. Note that the anyons have a nearest-neighbor repulsive interaction but an attractive interaction at next-nearest neighbor distance.}
	\label{fig:interaction}
\end{figure}
In the pure Kitaev model, described by $H_K$, a weak Zeeman term $\Delta H_h = \sum_i \vec h\cdot \vec \sigma_i$, induces $\kappa \approx \frac{h^3}{K^2}$. However, when generic symmetry-allowed spin interactions are present, $\kappa$ can arise at linear order in $h$ \cite{kasahara2}. In this work, we, therefore, consider $\kappa$ as an unknown, independent parameter.

{\em Ising anyons:} In the CSL phase, each vison carries a Majorana zero mode (MZM) exponentially localized around it. The localization length of the MZM is inversely proportional to the Majorana gap, $\xi \propto \Delta_m^{-1}$. 
When the separation $\vec r$, between two anyons $\sigma_1$ and $\sigma_2$, is larger than the localization length of the MZMs attached to them (i.e, $r \gg \xi$) the pair carries a non-local (complex) fermion mode, $\psi=\gamma_1+i \gamma_2$ at energy $\epsilon_0=0$, where $\gamma_k$ is the MZM localized around $\sigma_k$.

If $\psi$ is empty (occupied), the anyons are said to be in the bosonic (fermionic) sector. 
When $r$ is comparable to $\xi$, the fermionic mode obtains a non-zero energy, as the two MZMs begin to overlap. This results in an effective interaction between the two anyons that is exponentially small in $\vec r$, with an oscillatory pre-factor. In the vacuum sector, this interaction can be well-approximated by the analytical form \cite{topoliquid,lahtinennjp,Pachos_2012,Knollethesis}
\begin{align}
	V(\vec r) = V_0\left(1-e^{-\frac{r}{\xi}}\cos(\lambda \vec K_D \cdot \vec r)\right).
	\label{eq:interaction}
\end{align}

Here $\vec K_D = (\frac{4 \pi}{3\sqrt{3}},0)$ is the Dirac point in the first Brillouin zone, and  $V_0$, a function of $\Delta_m$, $\xi$ and $\lambda$ are fitting parameters obtained the numerically for a given value fo $\kappa$, as shown in Fig.~\ref{fig:interaction}.

The localized MZM also induces anyonic statistics for the visons. Transporting one anyon around another generally results in a {\em braiding} phase for the total wavefunction. This braiding phase depends on the sector to which the pair belongs. Exchanging the two anyons results in phase factors described by the following matrix elements.
\begin{align}
R^{\sigma \sigma}_{\mathbb 1} = e^{-\nu\frac{i\pi}{8}},\qquad R^{\sigma \sigma}_{\psi} = e^{\nu\frac{i3\pi}{8}}
\end{align}
where $R_c^{ab}$ denotes exchanging  particles $a$ and $b$ whose fusion outcome is $c$. Here, $\nu=\text{sgn}(\kappa)=\pm 1$ is the Chern number of the (matter) Majorana band, whose sign is determined by the external magnetic field direction. We use $\text{sgn}(\kappa)=1$, in our calculations.
For a bosonic anyon-pair, the braiding phase, corresponds to a double exchange (anti-clockwise) and is thus given by $\tilde{\theta}_\sigma=\pm 2\theta_\sigma=\pm\frac{\pi}{4}$. $\theta_\sigma$ is a universal property of the CSL phase which is described by an Ising topological field theory \cite{Kitaev06}. The anyons are, hence, also known as Ising anyons.

\section{Generic Kitaev Liquid and dynamical Anyons} 
\label{sec:GKSL}
\begin{figure}
	\centering
	\includegraphics[width=0.5\textwidth]{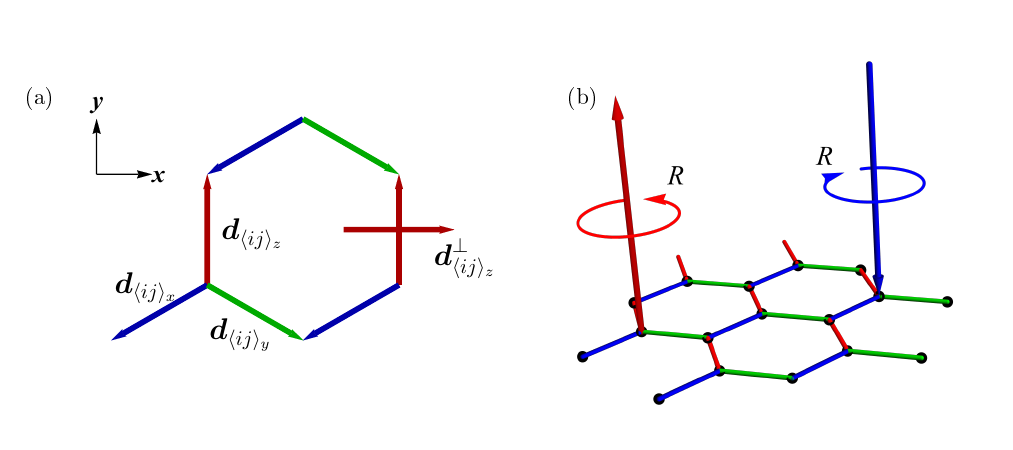}
	\caption{\textbf{Raman scattering geometry.} (a) Definitions of the vectors $d_{\nn{ij}_\alpha}$ and $d^\perp_{\nn{ij}_\alpha}$. (b) Incoming light  is incident perpendicular to the honeycomb plane and the light that is reflected back is detected. Shown here is a setup where the incident and reflected light waves are right-circularly polarized. This processes transfers a finite angular momentum, $\Delta S_{ph}=2\hbar$ from the photon to the system.}
	\label{fig:geometry}
\end{figure}
While $H_{CSL}$ is exactly solvable due to the static nature of the anyons, any material realization of the model will have additional interactions that break this integrability. Obviously the Zeeman term already affects the anyons at linear order in perturbation theory by hopping them from one plaquette to another. In the limit where such interactions are small, one can use a controlled perturbation theory to obtain the low-energy dynamics of the anyons. We thus define a generic  chiral Kitaev spin liquid by
\begin{align}
	H_{gCKSL} = H_{CSL}+ \Delta H
\end{align} 
where $\Delta H$ represents all possible symmetry-allowed exchange interactions.
The most relevant of such terms often include the off-diagonal symmetric exchange, $\Delta H_\Gamma =  \Gamma \sum_{\nn{ij}_\alpha}\sigma^\beta_i \sigma^\gamma_j+\sigma^\gamma_i \sigma^\beta$ with $\alpha \ne \beta \ne \gamma$, and the Heisenberg term $\Delta H_J = J\sum_{\nn{ij}} \vec \sigma_i \cdot \vec \sigma_j$, together forming the so-called `$J-K-\Gamma$' model  \cite{gammaRau,trebst2017kitaev}. The Zeeman field is also be treated as a perturbation within our model.

As a major consequence of integrability breaking, $[\hat W_p,\Delta H]\ne 0$, the gauge field becomes a dynamical degree of freedom, transforming the anyons into mobile quasiparticles. In Ref.~\cite{prxvison}, we have analyzed, using a controlled many-body perturbation theory, the dynamics of a single vison/anyon in a weakly perturbed Kitaev model (both in the gapless and gapped phases). It was found that the sign of the Kitaev coupling strongly influences the anyon band structure. In a ferromagnetic Kitaev model ($K<0$), a single Ising anyon obtains a tight-binding dispersion on a triangular lattice with a nearest neighbor ($nn$) hopping rate $t^{(1)} \propto h$, and a next-nearest neighbor ($nnn$) hopping rate $t^{(2)} \propto \Gamma$. In contrast, in the antiferromagnetic model ($K>0$), the anyon unit-cell is enlarged, due to projectively implemented translations, resulting in two bands with finite Chern numbers. The Heisenberg term induces hopping of single anyons only at higher orders in perturbation theory, but can induce large pair-hopping amplitudes \cite{Batista}. As most candidate materials are believed to posses a ferromagnetic Kitaev term, we will mainly focus on $K<0$ in this work and comment on the antiferromagnetic case towards the end.

\section{Light-Anyon coupling }

\label{sec:ramanop}
The general form of Raman operators for spin-orbit Mott insulators, relevant to Kitaev materials can be obtained by a symmetry analysis of the light-spin system. A more microscopic approach was recently used by Yang \textit{et al.} in Ref.~\cite{nonLF_perkins}, to derive the light-spin coupling beyond the conventional Loudon-Fleury mechanism. The total dipole Raman operator thus take the form

\begin{align}
	\mathcal R = \mathcal R^K + \mathcal R^H + \mathcal{R}^\Gamma+\mathcal{R}^h,
\end{align}
\begin{figure}[t!]
	\centering
	\includegraphics[width=.5\textwidth]{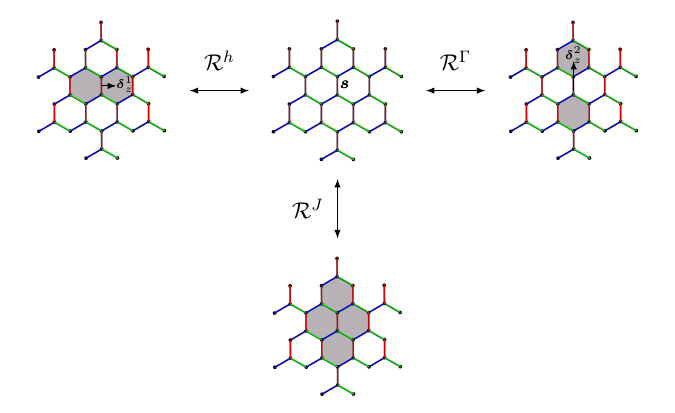}
	\caption{ Illustration of vison creation processes by each (local) Raman operators given by Eq.~\eqref{eq:raman_operators}, acting on a bond at position $\vec s$. $\mathcal R^h$ creates two visons at $\vec s\pm \vec \delta^{(1)}_z$ and $\mathcal R^\Gamma$ creates visons at $\vec s\pm \vec \delta^{(2)}_z$. $\mathcal R^H$ creates four adjacent anyons which requires a higher energy cost.}
	\label{fig:ramanOps}
\end{figure}
where 
\begin{align}
	\label{eq:raman_operators}
	\mathcal{R}^K=&\sum_{\langle ij\rangle_\alpha}{P}^K_{ij} \sigma^\alpha_i\sigma^\alpha_j, \nonumber \\
	\mathcal{R}^J=&\sum_{\langle ij\rangle_\alpha}{P}^J_{ij} \vec \sigma_i\cdot\vec \sigma_j,\nonumber \\
	\mathcal{R}^\Gamma=&\sum_{\langle ij\rangle_\alpha}{P}^\Gamma_{ij}(\sigma^\beta_i \sigma^\gamma_j+\sigma^\beta_j \sigma^\gamma_i)\nonumber \\
	\mathcal{R}^h=&\sum_{\langle ij\rangle_\alpha}{P}^h_{ij}(\sigma^\alpha_i+\sigma^\alpha_j).
\end{align}
 ${P}_{ij}(\vec e_{in},\vec e_{out})$ are polarization dependent pre-factors for a given link $\nn{ij}$, where $\vec e_{in}$ and $\vec e_{out}$ denote the polarization vectors of incident and reflected light respectively. The Raman operators includes only electric dipole transitions  induced by the electric field of the light, using that magnetic dipole transitions and multipole transitions are typically strongly suppressed (by powers of the fine-structure constant).
 
 The zero-temperature Raman response is thus given by the intensity
\begin{align}
	I(t) &= \bra{\Psi_0}\mathcal R(t)\mathcal R(0) \ket{\Psi_0} \nonumber \\&=  \bra{\Psi_0}e^{-i H t}\mathcal R e^{i H t}\mathcal R\ket{\Psi_0},
	\label{eq:raman_int1}
\end{align}
where  $\ket{\Psi_0}$ is the ground state of the gKSCSL. Within our model, all the excitations are gapped and hence the ground state with no visons (anyons), at the level of first order perturbation theory can be approximated by the ground state of the pure Kitaev model:
\begin{align}
\ket{\Psi_0} = \hat{P}\ket{0}\ket{M_0}
\end{align} where $\ket{M_0}$ denotes the ground state of the matter fermions  and $\ket{0}$ the vacuum of the gauge sector.  Here, $\hat P$ is an operator that projects the Majorana wavefunctions into the physical Hilbert space of the spins.
A general eigenstate of the unperturbed chiral Kitaev model (i.e, $\Delta H=0$) can thus be expressed as
\begin{align}
	\ket{\Psi_n(\{\vec r_i\} )} = \hat{P}\ket{\{\vec r_i\}}\ket{M_n\left(\{\vec r_i\}\right)}
	\label{eq:wf}
\end{align}
where the coordinates $\vec r_i$ denote the positions of the anyons and $\ket{M_n\left(\{\vec r_i\}\right)}$ corresponds to the eigenstates of the matter Majoranas in the presence of the visons.
\subsection{Anyon creation by Raman processes}

Let us examine the coupling between the Raman operator and the excitations of a gCKSL. As illustrated in Fig.~\ref{fig:ramanOps}, different Raman terms in Eq.~\eqref{eq:raman_operators} act on the ground state to excite multiple anyons locally. $\mathcal R^K$ is a Kitaev-like term which, to leading order in $\Delta H$, does not excite the visons but only couples to the matter fermions. Within our assumption, the Majoranas have a much larger energy gap compared to the visons, thus making the response from $\mathcal R^K$ appear only at higher frequencies $\omega \sim \Delta_m \gg E_\sigma^0$. The terms $\mathcal R^\Gamma$ and $\mathcal{R}^h$, however, create pairs of nearest-neighbor ($nn$) and next-nearest-neighbor ($nnn$) anyons respectively. The Heisenberg term, $\mathcal R^J,$ excites four adjacent visons, contributing only at higher frequencies $\omega\sim 4 E^0_\sigma$.  The low energy Raman response, hence, arises mainly from $\mathcal R^{2\sigma} = \mathcal R^\Gamma+\mathcal{R}^h$ induced pair-creation. The corresponding polarization dependent pre-factors are given by \cite{nonLF_perkins} (see Eq.~\eqref{eq:raman_operators}),
\begin{widetext}
	\begin{align}
			{P}_{\langle ij \rangle_\alpha}^\Gamma = -\frac{\Gamma_R}{2}\vec e_{in}\cdot \vec d_{\langle ij \rangle_\alpha} ~ \vec e^*_{out}\cdot \vec d_{\langle ij \rangle_\alpha},\quad
			{P}_{\langle ij \rangle_\alpha}^h = -i \frac{h_R}{2}(\vec e_{in}\cdot \vec d_{\langle ij \rangle_\alpha}^\perp  \vec e^*_{out}\cdot \vec d_{\langle ij \rangle_\alpha}-\vec e_{in}\cdot \vec d_{\langle ij \rangle_\alpha} \vec e^*_{out}\cdot \vec d_{\langle ij \rangle_\alpha}^\perp).
		\end{align}\label{eq:RamanVertices}
\end{widetext}
Vectors $\vec d_{\nn{ij}_\alpha}$ and $\vec d^\perp_{{\nn{ij}_\alpha}}=d^\perp_{{\nn{ij}_\alpha}}\cross \hat{\vec z}$ are defined in Fig.~\ref{fig:geometry}a. 
 $\vec e_{in}$ and $\vec e_{out}$ are the complex polarization vectors of the incoming and outgoing light respectively  (assumed here to be in the xy-plane).  
 
 In a typical Raman experiment, light is reflected on the surface of the sample. We use a convention where incoming light is oriented along the $-\hat z$ direction, while outgoing light is emitted in the $+\hat z$ direction. As we will particularly focus on circularly-polarized Raman scattering geometries, let us express the incoming and outgoing light in circular polarization basis as follows:
\begin{align}
\vec e_{in}^{R/L}&=e_{-/+}, \ \vec e_{out}^{R/L}=e_{+,-}, \ \text{with} \ \vec e_{\pm}&=\frac{1}{\sqrt{2}} \left(\begin{array}{c} 1 \\ \pm i \end{array}\right). 
 \end{align}
 Here, $R (L)$ denotes right(left)-circular polarization. While using circular polarized light, one can consider two types of scattering processes depending on the change in the spin part of the angular momentum of the incoming photon, denoted by $\Delta S_z$. The Raman intensities in different channels can be then denoted by
\begin{align}
I_{RL}(\omega)&=I_{--}(\omega), \quad I_{LR}(\omega)=I_{++}(\omega)  \quad \text{with } \Delta S^\text{ph}_z=0 \nonumber \\
I_{LL}(\omega)&=I_{+-}(\omega)  \quad \text{with } \Delta S^\text{ph}_z= -2 \hbar \nonumber \\
I_{RR}(\omega)&=I_{-+}(\omega)  \quad \text{with } \Delta S^\text{ph}_z= 2 \hbar
\end{align}
where the first  (second) index refers to outgoing (incoming) light. For left-handed (right-handed) light the photon spin is parallel (antiparallel) to  the orientation of the beam. Therefore, in a $RL$ or $LR$ process (which also describe ordinary reflections on a mirror) the spin-part of the angular momentum of the photon is not changed, $\Delta S^\text{ph}_z=0$, while in a $LL$ process, the angular momentum $\Delta S^\text{ph}_z= -2 \hbar$ is transferred to the sample.


Consider the creation of an anyon-pair out of vacuum by $\mathcal{R}^\Gamma$ on a $z$ bond $\nn{ij}$ with position $\vec s$ as shown in Fig.~\ref{fig:ramanOps}. The corresponding matrix elements evaluated using the representation given by Eq.~\eqref{eq:wf} expressed in a fixed the gauge, and implementing the projection operator (See Ref.~\cite{prxvison} for details) takes the form
\begin{widetext}
\begin{align}
	\bra{\Psi_n}\mathcal R^\Gamma (\vec s)\ket{\Psi_0}=
P^\Gamma_{ij}\bra{M_n\left(\vec s-\vec \delta_z^{(2)},\vec s +\vec \delta_z^{(2)}\right)}(-1 - ic_ic_j)\ket{M_0}.
\label{eq:me_gamma}
\end{align}

\end{widetext}
 Each term in $\mathcal R^\Gamma$ therefore couples the ground state and excited states with a $nnn$ vison pair with $n=0,2,4,...$ Majorana excitations.
 
$\mathcal R^\Gamma$ thus creates a pair of anyons plus an even (or zero) number of Majorana fermions. While the lowest energy excited state is given by $n=0$ which corresponds to a bosonic anyon-pair, at higher energies ($n>0$) one can occupy the localized fermion mode, resulting in a fermionic-anyon pair plus an extra high energy Majorana mode above the gap $\Delta_m$. Such states however only contribute to the Raman response at higher frequencies and are not considered here.
 
Similarly $\mathcal R^h(\vec s)$ creates a pair of $nn$ anyons with matrix elements of the form  
\begin{widetext}
	\begin{align}
		\bra{\Psi_n}\mathcal R^h (\vec R)\ket{\Psi_0}=
		P^h_{ij}\bra{M_n\left(\vec s-\vec \delta_z^{(1)},\vec s+\vec \delta_z^{(1)}\right)}(ic_i + ic_j)\ket{M_0}.
		\label{eq:me_h}
	\end{align}
\end{widetext}
In contrast to  Eq.~\eqref{eq:me_gamma}, here $\ket{M_n}$ corresponds to eigenstates with $n=1,3,5,...$ Majorana excitations. This also implies that the ground state of a nearest-neighbor anyon-pair must have the lowest lying Majorana mode occupied; $n=1$ \cite{joy2024gauge}. The lowest such mode corresponds to the localized fermion mode bound to the anyon pair which becomes a zero mode when the anyons are well-separated. Occupying this mode thus results in a bosonic-anyon pair excitation. As a side remark, the occupation of this mode due to the above parity constraint is responsible for the nearest-neighbor repulsive interaction between the anyons, as seen in Fig.~\ref{fig:interaction}. As mentioned in the case of $\mathcal R^\Gamma$, one may instead occupy a high energy fermion (above the Majorana gap $\Delta_m$) which would result in a fermionic-anyon pair plus a high energy Majorana - a process that only contributes at high frequencies. Thus we only consider the dynamics of the bosonic-anyon pair which is the relevant low-energy excitation in the limit $T\to 0$.
 
Transforming to the Lehman representation of Eq.~\eqref{eq:raman_int1}, we can introduce eigenstates of $H$ projected onto the bosonic anyon-pair sector, denoted by $\ket{\lambda}$. Fourier transforming to frequency space, we obtain the Fermi's golden rule
\begin{align}
	I(\omega) = \frac{2\pi}{\hslash}\sum_{\lambda}|\bra{\lambda}\mathcal R^{2\sigma}\ket{\Psi_0}|^2 \delta(\omega - (E_\lambda-E_0)).
	\label{eq:raman_intfreq}
\end{align}
where $\mathcal R^{2\sigma}=\mathcal R^h+\mathcal R^\Gamma$.
Our task now is to obtain the eigenstates $\ket{\lambda}$.
\section{two-anyon effective model}
\label{sec:effective}
  \begin{figure*}[t!]
	\centering
	\begin{tikzpicture}
		\node[anchor=center] (image) at (-7,0) { \includegraphics[width=0.6\textwidth]{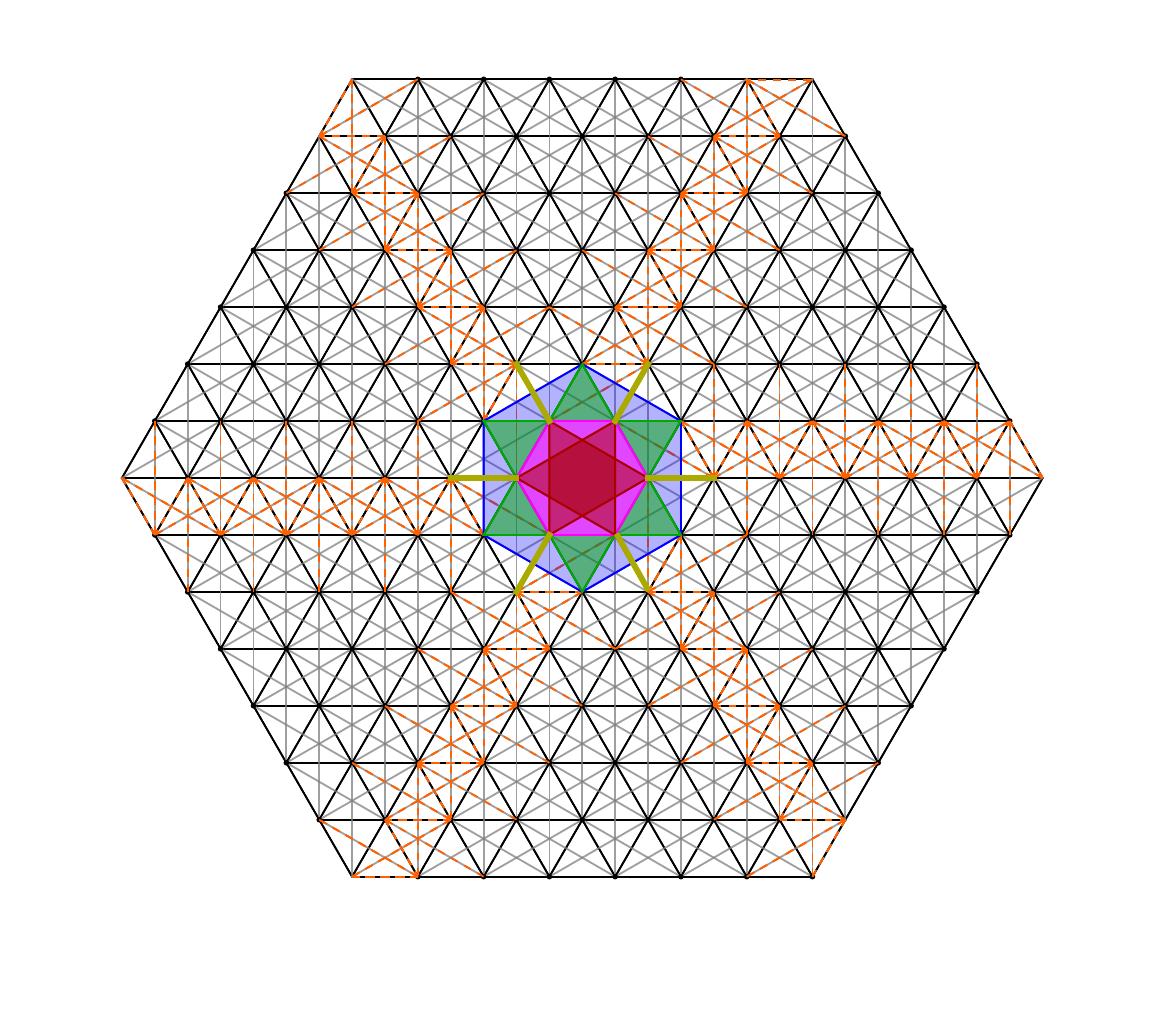}};
		\node[anchor=center] (text) at (-10.5,4.5) {\small{(a)}};
			\node[anchor=center] (image) at (2,-2) {\includegraphics[width=0.4\textwidth]{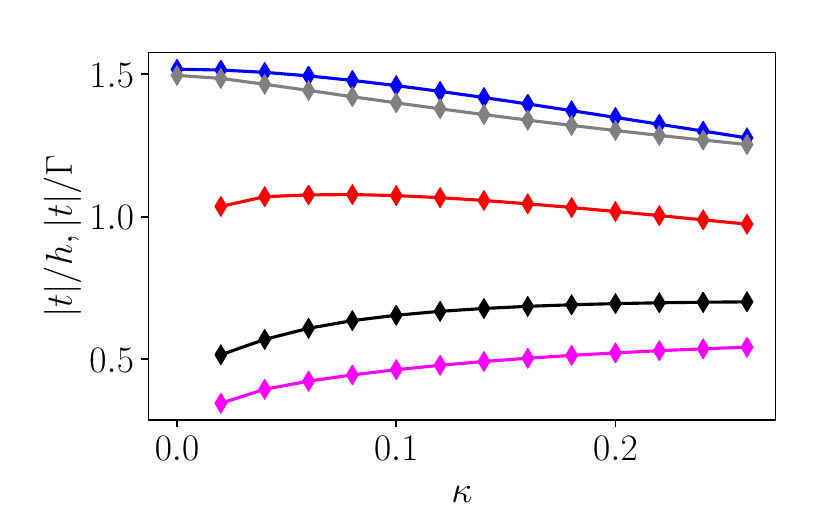}};
		\node[anchor=center] (image) at (2,2.5) {\includegraphics[width=0.4\textwidth]{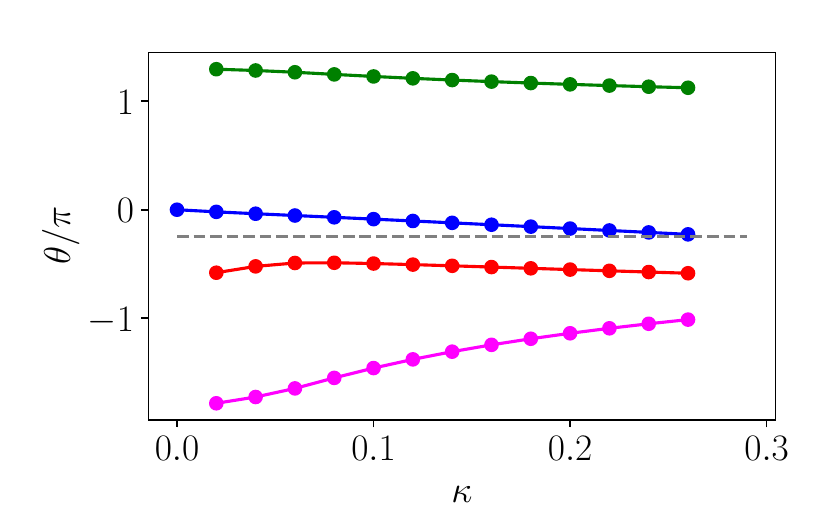}};
		\node[anchor=center] (text) at (-1,4.5) {\small{(b)}};
		\node[anchor=center] (text) at (-1,0) {\small{(c)}};
		\node[anchor=center] (text) at (4.7,3) {$\theta_{\textcolor{blue}{\hexagon}}$};
		\node[anchor=center] (text) at (4.7,4) {$\theta_{\textcolor{green}{\triangle}}$};
		\node[anchor=center] (text) at (4.7,2.5) {$\theta_{\textcolor{red}{\triangleleft}}$};
		\node[anchor=center] (text) at (4.7,2) {$\theta_{\textcolor{magenta}{\hexagon}}$};
		\node[anchor=center] (text) at (4.5,-.6) {$t_{\textcolor{blue}{\hexagon}}$};
		\node[anchor=center] (text) at (4.7,-1.2) {$t^{(2)}_{\sigma}$};
		\node[anchor=center] (text) at (4.7,-2) {$t^{(1)}_{\sigma}$};
		\node[anchor=center] (text) at (4.,-1.4) {$t_{\textcolor{red}{\triangleleft}}$};
		\node[anchor=center] (text) at (4.7,-3) {$t_{\textcolor{magenta}{\hexagon}}$};
	\end{tikzpicture}
	\caption{Effective lattice hopping model describing the relative motion of two Ising anyons in a generic Kitaev liquid. For $|\vec r|>\sqrt{3}$ (outside the blue hexagon), transporting one of two Ising anyons (created out of vacuum) anti-clockwise around the other results in a phase of $-\frac{\pi}{4}$. This is implemented by Peierls phases of $e^{-i\pi/24}$ multiplying the hopping amplitudes (orange dashed lines). We exclude hoppings to the origin, modeling a hard-core constraint for the anyons. This neglects a higher-order virtual process where the visons annihilate which is beyond the precision of our linear-order perturbation theory. When the anyons are close to each other, interaction effects lead to non-universal braiding phases dependent on $\kappa$. We calculate these phases and corresponding hopping rates microscopically within each hexagonal (or triangular) loops (drawn as colored polygons in (a)) up to a separation of $|\vec r|=\sqrt{3}$ (blue hexagon). We choose the gauge such that hopping on the yellow links carry a phase $\pi/24$ (along radially outward direction) in order to implement the mirror symmetry of the original honeycomb lattice. In (b), we plot the values of these phases along each closed loop shown in (a) (in colors) calculated explicitly on a honeycomb lattice model of linear size $L=34$. In (c), we plot the absolute values of the hopping matrix elements (normalized by the strength of the perturbation $h$ and $\Gamma$) corresponding to the sides of the colored polygons in (a), using the same colors.}
	\label{fig:lattice}
\end{figure*}
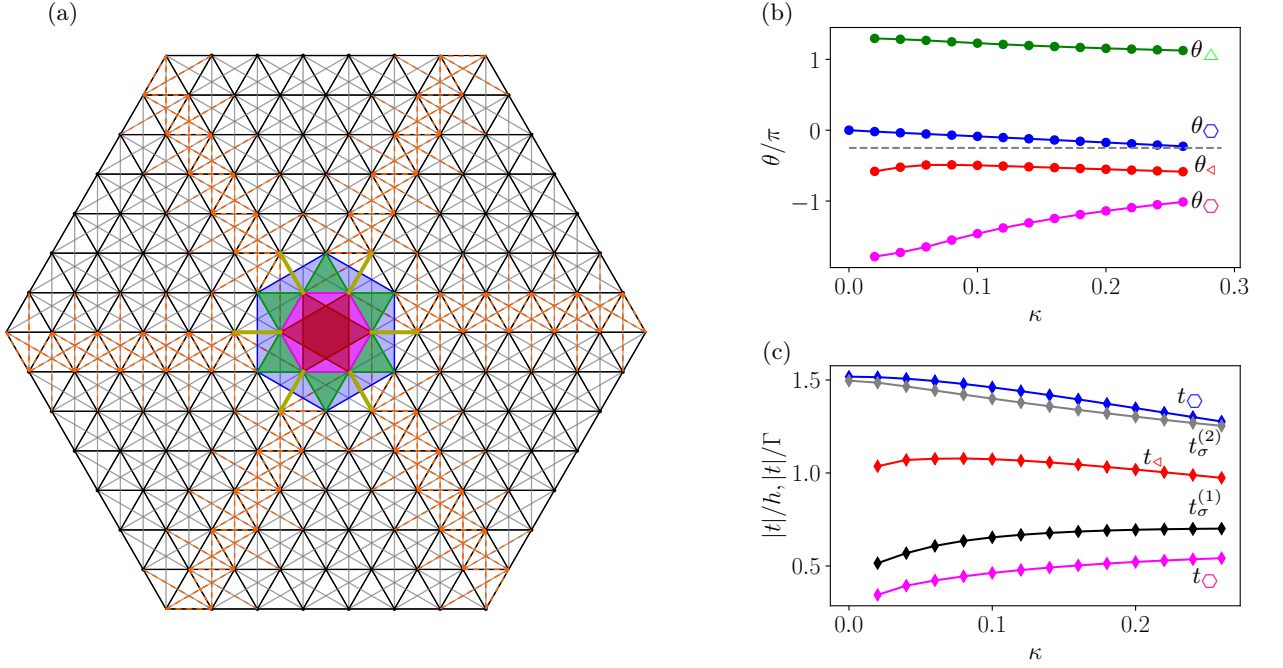
Due to their long-range statistical phase and short-range interaction effects, one cannot simply obtain the two-particle response from the single-particle dispersion. Instead, we will construct an effective lattice model to describe the dynamics of a pair of interacting Ising anyons in the limit where $\Delta H$ is weak.

The form of the two-particle Hamiltonian describing two anyons depends on the chosen gauge. Here, we denote the state of two anyons at positions $\vec r_1$ and $\vec r_2$ by $\ket{\vec r_1,\vec r_2} = \ket{\Psi_0(\vec r_1,\vec r_2)}$. We choose a bosonic gauge in which the particles are assumed to be bosons, i.e. $\ket{\vec r_1,\vec r_2}= \ket{\vec r_2,\vec r_1}$. To implement the non-trivial exchange phase, we attach a `magnetic' flux to each particle such that when one particle loops around another in an anti-clockwise fashion, it picks up a phase of $\tilde{\theta}_\sigma=-\pi/4$. The vector potential used to express the flux can be then implemented as Peierl's phases in the hopping amplitudes. The hamiltonian of the two particle system can be written in position space, on a triangular lattice as 

\begin{align}
	h_{2\sigma} = \sum_{\substack{\vec r_1, \vec r_2,\\ \vec r'_1,\vec r'_2}}t(\vec r_1,\vec r_2;\vec r'_1,\vec r'_2) \ket{\vec r'_1,\vec r'_2}\bra{\vec r_1, \vec r_2} +V(\vec r_1-\vec r_2)
	\label{eq:single_anyon_tb}
\end{align}
 The complex hopping amplitudes $t$ are directly related to the weak non-Kitaev interactions $\Delta H$, as discussed in Ref.~\cite{prxvison}. We restrict ourselves to $nn$ and $nnn$ hopping, denoted by $t^{(1)}$ and $t^{(2)}$ respectively. Longer range hopping processes arise only above $O(\Delta H^3)$ for the most relevant non-Kitaev interactions, such as the Heisenberg or $\Gamma$ couplings. Within our model, the hopping amplitudes are treated as unknown parameters.

Solving the Schrödinger equation involves decoupling the dynamics into the center of mass (COM) and relative coordinate frames.
Due to the overall translation symmetry, the total momentum $\vec K$ is a good quantum number and one can use the plane wave basis given by
\begin{align}
	\ket{\vec K,\vec r} = \frac{1}{L} \sum_{\vec r_1} e^{i\vec K \cdot (\vec r_1 + \frac{\vec r}{2})} \ket{\vec r_1,\vec r_1+\vec r},
\end{align}
where  $\vec r = \vec r_2-\vec r_1$ is the relative coordinate and $\vec K$ couples to the center of mass at $\vec R=\vec r_1+\vec r/2$. $L$ is the linear system size required for normalization. Calculating the matrix elements of $h_{2\sigma}$ in the relative coordinate basis leads to a tight-binding Hamiltonian on a triangular lattice formed by the relative coordinate vectors $\vec r$. The hopping amplitudes in this frame are given by
\begin{align}
 \tilde{t}^{(l)}_i(\vec r)= 2~{t}^{(l)}_i(\vec r)\cos(\frac{\vec K \cdot \vec \delta^{(l)}_i}{2})
 \end{align} 
 where $l=1,2$.
$\vec \delta^{(1)}_i$ and $\vec \delta^{(2)}_i$ denote the six $nn$ and $nnn$ vectors of a triangular lattice, defined in Fig.~\ref{fig:lattice}. Explicitly,
\begin{align} 
	&\vec \delta^{(1)}_1=(1,0),\,\vec \delta^{(1)}_2=(1/2,\sqrt{3}/2),\,\vec \delta^{(1)}_3=(-1/2,\sqrt{3}/2),\\ \nonumber
	&\vec \delta^{(1)}_4=-\vec \delta^{(1)}_1,\,\vec \delta^{(1)}_5=-\vec \delta^{(1)}_2,\,\vec \delta^{(1)}_6=-\vec \delta^{(1)}_3 \\ \nonumber
	&\vec \delta^{(2)}_i=\vec \delta^{(1)}_i+\vec \delta^{(1)}_{i+1}, \end{align}
	where $i$ is defined modulo 6.
As Raman scattering is a zero momentum transfer process, we will later set $\vec K=0$ in our calculations, leading to $\tilde{t}_i^{(l)}=2t^{(l)}_i$. Next, we will incorporate the effects of interaction between anyons into this lattice model. 

In the relative coordinate frame, the braiding phase $\tilde \theta_\sigma=-\frac{\pi}{4}$ (in the bosonic sector) between the two anyons can be implemented as an effective `magnetic' flux at the origin, $\vec r=0$. In Fig.~\ref{fig:lattice}a, we show how this is done in the triangular lattice by choosing a $C_6$ symmetric gauge where the the hopping amplitudes on the orange links are given by $|\tilde{t}^{(1,2)}|e^{-i\frac{\pi}{24}}$.

The universal braiding phase, $\tilde \theta_\sigma=-\frac{\pi}{4}$, however, is only valid when the anyons are well-separated, i.e, $r \gg \xi$.
When in proximity, the MZMs of two anyons start to overlap. This leads to a smeared out flux at the origin. We account for this by splitting the relative coordinate lattice into two regions.
For $|\vec r|\le \sqrt{3}$, we explicitly compute the hopping matrix elements and Berry phases as described in more detail below.  This is illustrated in Fig.~\ref{fig:lattice}a. For example, an $nn$ anyon pair taken around the origin (corresponds to the magenta loop in Fig.~\ref{fig:lattice}a) acquires a phase $\theta_{\textcolor{magenta}{\hexagon}}$ while an $nnn$-pair acquires a phase of $\theta_{\textcolor{blue}{\hexagon}}$. Fig.~\ref{fig:lattice}b  shows their values as a function of the Majorana gap $\Delta_m$. For a more accurate extrapolation of this effective `finite extend' of the flux, we also compute the flux through the triangles shaded in green. Beyond the the $nnn$ distance, we implement the universal exchange phase $\tilde{\theta_\sigma}$,  which is independent of $\Delta_m$ and is a universal property of the chiral spin liquid, shown by the dashed line in Fig.~\ref{fig:lattice}b. For large $\kappa$, $\theta_{\textcolor{blue}{\hexagon}}$ approaches the universal value, as expected.

The effective Hamiltonian of a pair of Ising anyons in the relative coordinate frame, for a given $\vec K$, can be thus written as (suppressing the $\vec K$ label)
\begin{widetext}
\begin{align}
	h^{rel}_{2\sigma} = 2\Delta^0_{\sigma} + {2}\sum_{\vec r\ne 0, i}\left(\tilde{t}^{(1)}_i\left(\vec r\right) \ket{\vec r + \vec \delta_i^{(1)}}\bra{\vec r}+ \tilde{t}^{(2)}_i\left(\vec r\right)\ket{\vec r+\vec \delta_i^{(2)}}\bra{\vec r} +\text{h.c}\right) + \sum_{\vec r\ne 0}V(\vec r).
\end{align}
\end{widetext}
where the site $\vec r= 0$ is excluded as two visons cannot occupy the same site.  The constant  $\Delta^0_{\sigma}$ is used to shift the energy of a single anyon.

 In our calculations, we start by fixing the value of the Majorana gap $\Delta_m$. The parameters $\xi, \lambda$ and $V_0$ that determine the interaction term $V(\vec r)$ are then obtained by fitting Eq.~\eqref{eq:interaction} to the numerical data as shown in Fig.~\ref{fig:interaction}. The hopping amplitude of a $nnn$ pair is approximately equal to that of a single anyon as shown in Fig.~\ref{fig:lattice}c. However, the value of an  $nn$ anyon-pair hopping amplitude is smaller than that of a single anyon by a factor of approximately $0.7$. Our model therefore has three free parameters for a fixed value of $\kappa$: $h,\, \Gamma$ and $\Delta^0_\sigma$. We parameterize the hopping amplitudes as $h=\frac{t}{2} \cos{(\phi)}$ and $\Gamma=\frac{t}{2} \sin(\phi)$. Thus in the effective lattice shown in Fig.~\ref{fig:lattice}, the red links correspond to hopping amplitudes with magnitudes $|\tilde t^{(2)}| = 2\Gamma t_{\textcolor{red}{\triangleleft}}$, the magenta links correspond to $|\tilde t^{(1)}| = 2h t_{\textcolor{magenta}{\hexagon}}$ and the blue links correspond to $|\tilde t^{(2)}| = 2\Gamma t_{\textcolor{blue}{\hexagon}}$. All $nn$ and $nnn$ hopping amplitudes outside the blue ring are given by $|\tilde t^{(1)}| = 2ht^{(1)}_{\sigma}$ and $|\tilde t^{(2)}| = 2\Gamma t^{(2)}_{\sigma}$ respectively. The Berry phases and the universal braiding phase are implemented using Peierls substitution.

Diagonalizing $h^{rel}_{2\sigma}$ on a finite lattice of linear dimension $L \sim 10^2$, we obtain the two-anyon energy $E^{2\sigma}_{n,\vec K=0}$ (measured relative to the ground-state energy) and eigenfunctions $\ket{\phi^{2\sigma}_n}$ written as
\begin{align}
	\ket{\Phi^{2\sigma}_n(\vec K=0)} = \sum_{\vec r} \chi_{n}(\vec r)\ket{0, \vec r}.
\end{align}
Note that due to our choice of a bosonic-gauge, only symmetric wavefunctions $\chi_{n}(\vec r)=\chi_{n}(-\vec r)$ are allowed. Numerically, we use rotation invariance by $60$ degrees to block-diagonalize the Hamiltonian. Due to the symmetry of the wavefunctions, only angular momentum subsectors with $l=0,2,4$ are relevant.

 \begin{figure*}
	\centering
	\begin{tikzpicture}
		\node[anchor=center] (image) at (-7,0) {\includegraphics[width=0.45\textwidth]{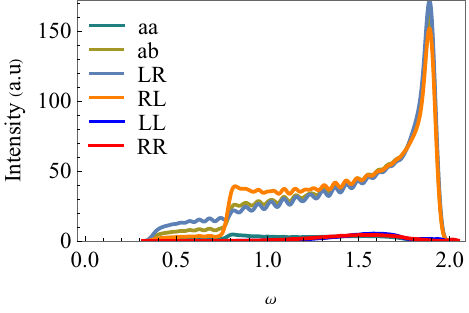}};
		\node[anchor=center] (text) at (-11.,2.5) {{(a)}};
		\node[anchor=center] (image) at (2,0) {\includegraphics[width=0.45\textwidth]{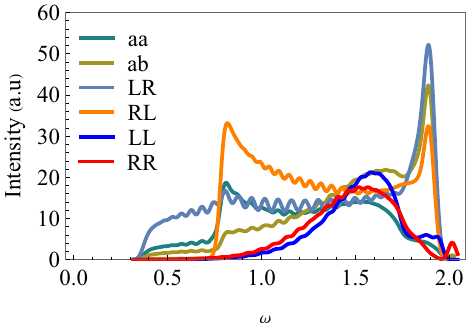}};
		\node[anchor=center] (image) at (-6.,1.3) {\includegraphics[width=0.17\textwidth]{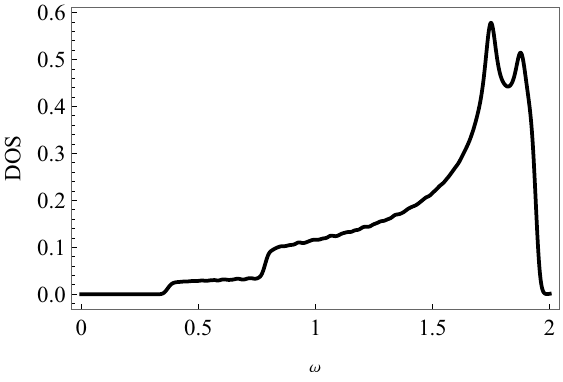}};
		\node[anchor=center] (text) at (-2.,2.5) {{(b)}};
		\node[anchor=center] (image) at (-7,-5.5) {\includegraphics[width=0.45\textwidth]{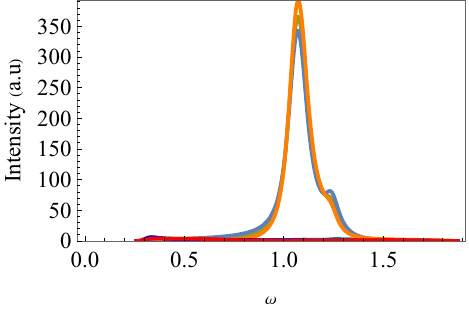}};
		\node[anchor=center] (text) at (-11.,-3) {{(c)}};
		\node[anchor=center] (image) at (2.2,-5.5) {\includegraphics[width=0.45\textwidth]{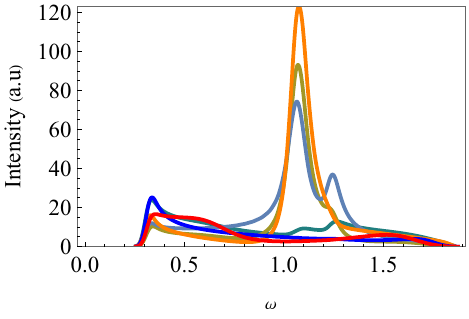}};
		\node[anchor=center] (image) at (-8,-4.3) {\includegraphics[width=0.17\textwidth]{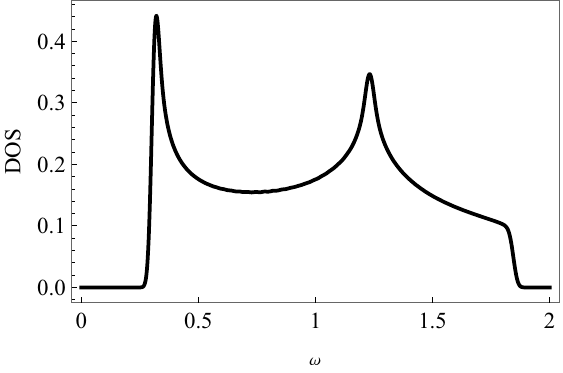}};
		\node[anchor=center] (text) at (-2.,-3) {{(d)}};
\end{tikzpicture}
	\caption{Low-energy continuum Raman response in the hopping dominated regime for different polarization of the light. $aa$ and $ab$ refers to linearly polarized light, while $L$ and $R$ refer to circularly polarization of light, see text. 
 Model parameters are $K=1, \kappa = 0.2$, with $\phi = 0.4\pi, 2\Delta^0_{\sigma}=1.5,\, t=-0.15$ for the top panels, and  $\phi = -0.1\pi ,2\Delta^0_{\sigma}=1.0, t=-0.18$ for the lower panels. Raman vertex parameters are $\Gamma_h/h_R=2$ for (a,c) and  $\Gamma_h/h_R=0.5$ for (b, d). 
The Raman signals show a strong dependence on the polarization of both the incoming and outgoing light. Their frequency dependence approximately trace the two-particle density of states shown in the insets of panels {a} and {c}. Furthermore, for panels a and b, anti-bound states are visible above the continuum, $\omega \approx 1.9$ and $\omega \approx 1.95$, in the $RR$ and $LL$ polarizations (and very weakly for $aa$ and $ab$ polarization) but not for the $RL$ and $LR$ channels due to angular momentum selection rules, see text. Broadening: $\Delta E=0.02$ is used. Small oscillations in the signal arise from finite-size effects.
}\label{fig:continuum} 
\end{figure*}
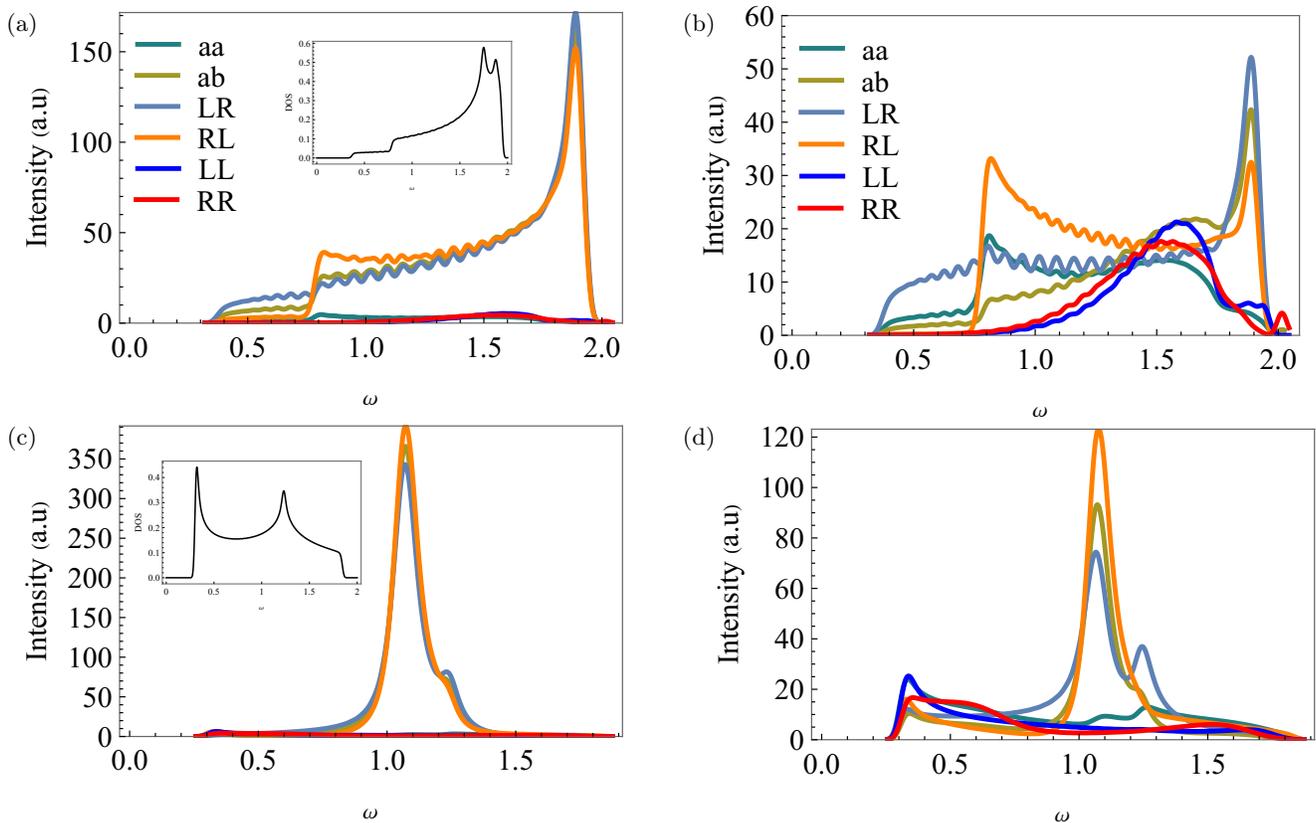

\section{Raman spectrum}
\label{sec:spectrum}
Substituting the obtained eigenstates into Eq.~\eqref{eq:raman_intfreq} and carrying out the summations, we can obtain a concise expression for the Raman intensity
\begin{align}\label{eq:Iw}
	I(\omega) = \frac{2\pi}{\hslash}\sum_n |M^h_n + M^\Gamma_n|^2 \delta(\omega-E^{2\sigma}_{0,n}),
\end{align}
where we define the matrix elements
\begin{align}
	M^h_n =& \sum_j \chi^*_n\left(\vec \delta^{(1)}_j\right)\bra{\Psi_0\left(0,\vec \delta^{(1)}_j\right)}\mathcal R^h\left({\vec \delta^{(1)}_j}\right)\ket{\Psi_0}, \nonumber \\
	M^\Gamma_n =& \sum_j \chi^*_n\left(\vec \delta^{(2)}_j\right)\bra{\Psi_0\left(0,\vec \delta^{(2)}_j\right)}\mathcal R^\Gamma\left(\vec \delta^{(2)}_j\right)\ket{\Psi_0}
	\label{eq:raman_me}
\end{align}

where the index $j=1,2,..,6$ enumerates six different $nn$ (or $nnn$) bonds around the origin. For our plots, we broaden the $\delta$ functions using Gaussians, $\delta(\omega) \to \frac{e^{-(\omega/\Delta E)^2/2}}{\sqrt{2\pi} \Delta E}$. The matrix elements can be evaluated numerically for a finite size system using the methods discussed in Ref.~\cite{prxvison,Batista,inti} (also see Appendix \ref{appA}). The center-of-mass momenta $\vec K$ does not appear in the response as Raman scattering is a zero momentum transfer process.

The $C_6$ rotational symmetry of the lattice can be employed to write the Raman matrix elements in the angular momentum eigenbasis labeled by $l$. For example, in the $RL$ and $LR$ channels (within our bosonic gauge) the matrix elements of $\mathcal R^h$ can be written in $l=0$ basis. However, for $LL$ $(RR)$ channels, matrix elements of $\mathcal R^\Gamma$ can be expanded in the $l=\pm 2$ basis.
\begin{align}
	\bra{\Psi_0\left(0,\vec \delta_j\right)}\mathcal R^{2\sigma}_{LR/RL} \left({\vec \delta_j}\right)\ket{\Psi_0} &\sim |R^\Gamma| \pm |R^h| \nonumber \\
	\bra{\Psi_0\left(0,\vec \delta_j\right)}\mathcal R^{2\sigma}_{LL/RR} \left({\vec \delta_j}\right)\ket{\Psi_0} &\sim |R^\Gamma| e^{\pm i2 \phi_j}
	\label{eq:polarization}
\end{align}
where $R^\Gamma$ and $R^h$ are the amplitudes of the corresponding matrix elements and $\vec \delta_j = \left(\cos(\phi_j),\sin(\phi_j)\right)$.

Below, we discuss three main features of the Raman response: the continuum response, the onset of the signal at low and high frequencies and the appearance of (anti-) bound states.

\subsection{Continuum response and polarization dependence}
When the Majorana gap is large, the MZM overlap is strongly suppressed and the short range interaction is negligible when compared to the hopping amplitudes. The response therefore display the two-anyon continuum. In Fig.~\ref{fig:continuum}, we show the Raman spectra in different polarization channels for different values of the hopping parameters and Raman vertices, for $\kappa=0.2 \,K$.

For most parameters, we obtain a strong dependence of the Raman response on the polarization channels. 
Remarkably, for circularly polarized light, the Raman intensity depends strongly on the polarization of the incoming photon: the amplitudes of $I_{RR}(\omega)$ and $I_{LL}(\omega)$ can be completely different. 
Also the $LR$ and the $RL$ response take different values. This has to be compared with time-reversal and inversion symmetric systems, where $I_{LR}=I_{RL}$ and $I_{LL}=I_{RR}$.
There are two origins of the polarization dependence: Raman vertices and the anyonic pair-wavefunctions. Due to angular momentum selection rules, the Raman matrix element $M^\Gamma_n$ is identical for $LR$ and $RL$ processes while $M^h$ changes sign (see Eq.~\eqref{eq:polarization}). Thus, there is either a constructive or destructive interference of the two vertices depending on the polarization.
In contrast, for $LL$ and $RR$ channels, only  $\mathcal R^\Gamma$ contributes, while the polarization dependent pre-factors $\mathcal P_{\nn{ij}}^h$ of $\mathcal R^h$ evaluates to zero. In this case, the finite values of $I_{LL}-I_{RR}$ arise purely from the anyonic wave functions and therefore from the complex hopping matrix elements in our effective Hamiltonian that encode the exchange statistics and Berry phases of the anyon-pairs. This observation is consistent with the general notion that cross-circularly polarized light scattering can be used to probe to time-reversal breaking orders \cite{mcd,gravitons}

A comparison of the left and right panels in Fig.~\ref{fig:continuum} shows the influence of Raman vertices on the relative size of the peaks in the signal. A large part of the signal approximately traces features (van Hove singularities and jumps due to band minima/maxima) in the two-particle density of states (at $\vec K=0$) shown in the insets of Fig.~\ref{fig:continuum}a and c. 
These features depend sensitively on the sign and relative size of nearest-neighbor and next-nearest neighbor hopping of the anyons, see top and bottom panels of  Fig.~\ref{fig:continuum}.
Furthermore, there can be anyonic bound states and anti-bound states discussed in Sec.~\ref{sec:bound} below.

\subsection{Power-law onset of the signal}
 \begin{figure*}[t]
	\centering
	\begin{tikzpicture}
		\node[anchor=center] (image) at (-7,0) {\includegraphics[width=0.44\textwidth]{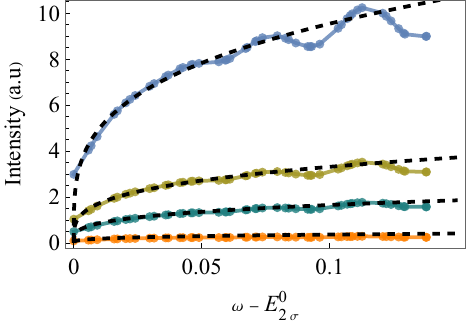}};
		\node[anchor=center] (text) at (-11.,3) {{(a)}};
		\node[anchor=center] (image) at (2,0) {\includegraphics[width=0.45\textwidth]{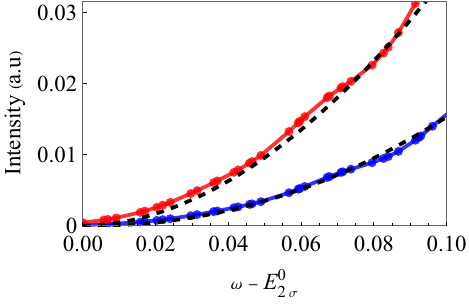}};
		\node[anchor=center] (text) at (-2.,3) {{(b)}};
	\end{tikzpicture}
	\caption[Power-law onset and polarization dependence]{Power-law onset of the Raman signal for model parameters corresponding to that of Fig.~\ref{fig:continuum}b. In linear ($a a$ and $a b$) and parallel-circular ($RL$ and $LR$) polarization channels, the signal onset shows the power-law $I(\omega)\sim A (\omega-E^0_{2\sigma})^\frac{1}{8}+B(\omega-E^0_{2\sigma})^\frac{1}{4}$ (dashed lines are the fits with $A$ and $B$ as parameters), arising from the topological spin $\theta_\sigma = \frac{1}{8}$ of the Ising anyons. (b) The onset in the cross-circularly polarized channels exhibit a power-law $I(\omega)\sim C(\omega-E^0_{2\sigma})^{2\pm\frac{1}{8}}$ (dashed lines) for the $LL$ and $RR$ channels respectively. Weak oscillations in the signals arise from finite size effects. Color code is identical to that of Fig.~\ref{fig:continuum}. (broadening: $\Delta E = 0.013$). See main text for details.}\label{fig:power_law} 
\end{figure*}
The effects of the anyonic statistics is most directly visible in the spectral function at the onset of the signal. At these energies, anyon states at the bottom of the band contribute. Small momenta correspond to large separation of the anyons, where the only remaining interaction arises from mutual exchange statistics.

In a notable work, Morampudi \textit{et al.} \cite{morampudi} argued that long-range statistical interaction between anyons results in a characteristic power-law onset in all correlation functions that involves the creation of two anyons. Below, we reproduce the calculation of Ref.~\cite{morampudi} and refine it to include subleading corrections and, furthermore, in Sec. VI C, extend the theory to situations where anyons have multiple band minima.

The frequency dependence of $I(\omega)$, Eq.~\eqref{eq:Iw}, arises from two factors: the density of states $\sum_n \delta(\omega-E_{n}^{2 \sigma})$ and the matrix elements  $|M^h_n + M^\Gamma_n|^2$ . Ignoring bound states (discussed in the next section), the density of states at low energy shows a jump at the onset
\begin{align}
\int\frac{d^2 k}{(2 \pi)^2} \delta\!\left(\omega-(E^\sigma_{\vec k}+E^\sigma_{-\vec k})\right) \sim \theta(\omega-E_{2\sigma}^0)
\end{align}
where $E^\sigma_{\vec k}$ is the single-anyon dispersion. Here, we use that for two-particle scattering states,  $E_{n}^{2 \sigma}=E^\sigma_{\vec k}+E^\sigma_{-\vec k}$ in the thermodynamic limit.
To compute the matrix elements, we need the value of the two-particle wave function for low energies at short distances, $\vec \delta_j^{(1,2)}$, see Eq.~\eqref{eq:raman_me}. As discussed above, the statistical interaction  is encoded by a flux of strength $\alpha$ (in units of the flux quantum. $\alpha=0,1$ for bosons and fermions respectively) located at the center of the relative coordinate space.  Here,  $\alpha=1/8$ for our bosonic anyons.
To calculate both the leading and subleading low-$\omega$ behavior, we consider the relative-coordinates Schr\"odinger equation at low energies
\begin{align}
-\frac{1}{2 m^*} 
\left( 
 \frac{1}{r} \frac{\partial}{\partial r}\left(r  \frac{\partial}{\partial r}\right)-\frac{(l-\alpha)^2}{r^2} \right)  \Psi_l(r)=E \Psi_l(r), \label{eq:schroedingerLow}
\end{align}
where the boundary condition $\Psi_l(a)=0$ takes into account the effect of a hard-core potential, where $a$ is the lattice constant. Here
$l$ is the angular momentum quantum number. As we use a bosonic gauge, only symmetric pair-wavefunctions with an {\em even} $l$ are allowed, $l=2 n$ (this is only valid for a band minimum at the $\Gamma$ point as discussed below). The equation is solved by $\Psi_l(r) \sim \sqrt{\frac{k}{2 L}}(
 J_{|l-\alpha|}(kr)+ c(k) Y_{|l-\alpha|}(kr))$, with $E=k^2/(2 m^*)$ and $k>0$ with Bessel functions $J$ and $Y$ of the first and second kind. At short distances the asymptotic form of the Bessel function are given by
 \begin{align}  J_{\nu}(kr)\sim (kr)^{\nu}\end{align}
 	 and  
 	 \begin{align}
 	 	Y_{\nu}(kr)\sim -\frac{ \cos(\nu \pi)\Gamma[-\nu] (kr)^{\nu} }{2^{\nu}~\pi} - \frac{2^\nu \Gamma[\nu] (kr)^{-\nu}}{\pi} 
 	 \end{align} 
where $\Gamma[\nu]$ is the gamma-function.

Imposing the hard-wall boundary condition gives gives $c(k)=-Y_{|l+\alpha|}(ka)/ J_{|l+\alpha|}(ka)$. 
Using that $\sum_k \frac{k}{2 L}\dots=\int \frac{k d k}{2 \pi}\dots $ for large $L$, we can absorb the normalization of $\Psi_l(r)$ into the integration measure. Using that $k^2 \sim \omega - E_{2 \sigma}^0$, we obtain for our bosonic Ising anyon-pair in $LR$ and $RL$ Raman channels
\begin{align}
I_{LR/RL}(E_{2 \sigma}^0+\Delta \omega) \sim \Theta( \Delta \omega) \big[ ( \Delta \omega)^{1/8} &+c ( \Delta \omega)^{1/4}\\ \nonumber &+O(\Delta \omega^{3/8})\big],
\end{align}
where $c$ is a constant.

Note that the sub-leading terms are only weakly suppressed. Technically, it arises from the hardcore constraint and the asymptotic behavior of $Y_\alpha$ but also other short-ranged interaction terms
are expected to contribute.

In contrast, for $LL$ and $RR$ Raman processes, which change the spin angular momentum of the photon, only terms with $l=\pm 2$ contribute and we find
\begin{align}
I_{LL}(E^{2 \sigma}_0+\Delta \omega) \sim \Theta( \Delta \omega) ( \Delta \omega)^{2 + 1/8} ,\nonumber \\
I_{RR}(E^{2 \sigma}_0+\Delta \omega) \sim \Theta( \Delta \omega) ( \Delta \omega)^{2 - 1/8}.
\end{align}
 Thus, the suppression of intensity at the threshold energy is much larger in this case.

The analytic arguments are fully confirmed by our numerical calculation of the Raman response, as shown in Fig.~\ref{fig:power_law}a and b where dashed line have been fitted to the theoretically expected asymptotics.

\subsection{Power-law onset in systems with multiple band minima}

An interesting scenario arises when the single particle band has multiple minima located at $\vec k =\vec k^{\text{min}}_n$, $n=1,\dots, N^{\text{min}}$. These multiple minima (or `valleys') also describe the two-particle Hamiltonian in relative coordinates. The two-particle interactions couple the different minima. If we consider, however, the low-energy limit, the wave function has less and less weight close to the origin and scattering between minima is suppressed. Thus, we can approximate the wave function by a product of a valley wave function varying on a length scale set by $2 \pi/|k^{\text{min}}_n|$ and a smooth wave function  $\Psi_l(\vec r) = e^{i l \phi} \tilde \Psi_l(r)$, where  $\tilde \Psi_l(r)\sim\sqrt{\frac{k}{2 L}} J_{|l-\alpha|}(kr)$ solves the effective Schr\"odinger equation, Eq.~\eqref{eq:schroedingerLow}, for the statistical parameter $\alpha$.
An eigenstate of total angular momentum $l$ can thus be written approximately as 
\begin{align}
\chi_l(\vec r)&\approx  \Psi_{l-m}(\vec r) \frac{1}{\sqrt{N^{\text{min}}}} \sum_n e^{i \vec k^{\text{min}}_n \vec r} 
e^{i \frac{2 \pi}{6} m n} \nonumber \\
&= \Psi_{l-m}(\vec r) \Psi^v_m(\vec r) ,
\end{align}
with integer values of $m$, where $\Psi^v_m(\vec r)$ is the wave function of the valley degree of freedom, carrying angular momentum $m$.
The total wave function has to be symmetric in our bosonic gauge, $\Phi(\vec r)=\Phi(-\vec r)$, and thus $l$ has to be even while $m$ can take values $m=0,1,\dots 5$.

For the evaluation of the Raman matrix elements, Eq.~\eqref{eq:raman_me}, we have to evaluate the wave function at the positions $\vec \delta_j^{(1,2)}$. We distinguish three cases.
First, in the limit $\vec k^{\text{min}}_n\to 0 $, we should recover the result of the previous section, where a single band minimum was discussed. Indeed, we find in this limit that $\Psi^v_m(\vec \delta_j^{(1,2)}) =0$ for $m\neq 0$ and we recover $\chi_l(\vec r)\approx \Psi_l(\vec r)$, resulting in 
\begin{align}
I_{LR/RL}\sim (\Delta \omega)^{|\alpha|}, \qquad I_{LL/RR}\sim (\Delta \omega)^{2\pm \alpha}.
\end{align}

Second, we consider the case, where the band minima are located at the $K$ and $K'$ points (Brillouin zone corners), thus there are effectively only two band minima. In this case, $\Psi^v_m(\vec \delta_j^{(1)}) =0$ for $m\neq 0, 3$, while $\Psi^v_m(\vec \delta_j^{(2)}) =0$ for $m\neq 0$. This implies that the valley wave function can absorb the angular momentum $m=3$ for the $\mathcal R^h$ Raman operator, but not for the 
$\mathcal R^\Gamma$ operator. As $e^{i 3\times  2 \pi/6}=-1$, $\Psi^v_{m=3}(\vec r)$ is an odd function of $\vec r$ at $m=3$. This implies that both odd and even $\Psi_{l-m}(\vec r)$ are allowed for  $\mathcal R^h$ processes.
An alternative interpretation of this result is that due to the two band minima, we can treat the two particles effectively as distinguishable. We now obtain
\begin{align}
I_{LR/RL}\sim (\Delta \omega)^{|\alpha|}, \qquad I_{LL/RR}\sim (\Delta \omega)^{1\mp \alpha}.\label{eq:twoBandMin}
\end{align}
For a processes where the angular momentum of the photon is changed by $\pm 2$, we expect that $m=\pm 3$ can be absorbed by the valley part of the wave function. Within our model, however, the pre-factor of the  $(\Delta \omega)^{1\pm \alpha}$ vanishes as only $\mathcal R^\Gamma$ contributes to the $LL/RR$ processes, which does not couple to the valley-degree of freedom, as discussed above. Thus, we obtain, within our model, $I_{LL/RR}\sim (\Delta \omega)^{2\pm \alpha}$ for band-minima at $K$, $K'$. In a generic experiment, however,  $I_{LL/RR}$ might be more singular and we expect the Eq.~\eqref{eq:twoBandMin} to hold.

In the third case, when six band minima are located neither at the $K$, $K'$ points nor at the $\Gamma$ point, the valley part of the wave function can absorb arbitrary angular momenta, $m=0,1,\dots 5$. Thus, we find in this case as leading behavior
\begin{align}
I_{LR/RL}\sim  I_{LL/RR}\sim (\Delta \omega)^{\text{min}(|\alpha|,1-|\alpha|)}. 
\end{align}
In our case $\text{min}(|\alpha|,1-|\alpha|)=|\alpha|=1/8$ but the more general formula can be applied also for anyons with $|\alpha|>1/2$.

\subsection{Bound states}\label{sec:bound}
The short range interaction, due to its oscillatory nature produces a repulsive nearest-neighbour interaction and an attractive next-nearest neighbor interaction (see Fig.~\ref{fig:interaction}). This may induce bound states (anti-bound states) below (above) the continuum in certain parameter regimes. These states contribute sharp peaks to the Raman spectrum.

 \begin{figure}
	\centering
	\begin{tikzpicture}
		\node[anchor=center] (image) at (0,0) {\includegraphics[width=0.45\textwidth]{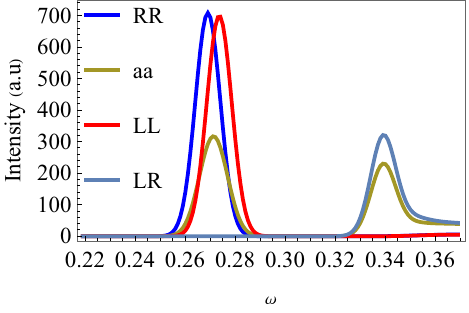}};
	\end{tikzpicture}
	\caption{For weak anyonic hopping rates, bound states of anyon pairs emerge due to a local minimum in the potential $V(\vec r)$ at short distances. The bound state are visible as sharp peaks in the Raman spectrum below the 2-anyon continuum (not shown here). Due to angular momentum selection rules (see main text), $I_{LL}$ can measure only $l=2$ bound states, $I_{RR}$ only $l=-2$, while  $I_{LR}$ and $I_{RL}$ only $l=0$ bound states. In linearly polarized light, all three bound states are visible simultaneously.
	Model parameters: $\kappa = 0.1 \,K, ~2\Delta_{\sigma} = 0.8 \,K, ~t=-0.08, ~\phi = -0.6 \pi, ~\Gamma_h/h_R=2, \text{broadening}: \Delta E=0.005$.}
	\label{fig:boundstates}
\end{figure}

For the parameters used in Fig.~\ref{fig:boundstates}, we find the emergence of three bound states, each labeled by their angular momentum quantum numbers $l=-2,0,+2$, which show specific polarization dependence. While the linear polarization channels couple to all the three bound states, parallel circularly polarization channels, $LR$ and $RL$, couple only to the $l=0$ states due to angular momentum selection rules.  However, $LL$ ($RR$) channels can be used to detect $l= +2 \,(-2)$ states.

The bound states arise from a shallow local minimum of the anyon-anyon potential, when the two anyons
are on next-nearest neighbor sites (with distance $\sqrt{3}$). These sites form a hexagon.
In the limit of small anyonic hopping, the energies of the bound states arise from the hopping on this hexagon and are approximately given by 
\begin{align} 
e^B_{l}\approx V_{nnn}+4 t_{\text{eff}} \cos\!\left(2 \pi \frac{l-\theta_{\text{eff}}}{6}\right)
\end{align}
where $t_{\text{eff}}$ and $\theta_{\text{eff}}$ are the effective hopping and effective flux for the hopping along the hexagon. Only $l=0, \pm 2$ are allowed by our selection rules.  For the parameters of  Fig.~\ref{fig:boundstates},  $t_{\text{eff}}>0$ (as $t_{nnn}>0$) and the $l=0$ bound state has the highest energy.  The splitting of the peaks in $I_{LL}$ and $I_{RR}$ arises from $\theta_{\text{eff}}$.

\section{AFM Kitaev models}
The analysis so far applies only to the FM Kitaev, where $K>0$. As mentioned in the introduction, this sign of the interaction is believed to be present in materials like $\alpha$-RuCl$_3$ \cite{ybkim,PhysRevLett.132.246702}.
Nevertheless, it is interesting to consider also  antiferromagnetic models,  $K<0$.
Remarkably, the Raman response changes drastically in this case. When one looks at the spectrum of the pure Kitaev model, the solutions for $K>0$ and $K<0$ are identical but the unitary transformation which  map the two models onto each other breaks translational invariance as has been pointed out by Kitaev \cite{Kitaev06}. 

For the ground state, 
one can absorb the sign of $K$ by multiplying all matter Majorana operators on one of the two sublattices by $-1$ while leaving the gauge links $u_{ij}$ unchanged. This harmless looking change implies, however, that the two models realize different types of projective symmetries as has been pointed out by Chen, Rao and Sodeman \cite{inti}: different types of gauge transformations have to be used to implement the translational symmetries of the two model. As projective symmetries can be used to classify the phases of spin liquids, one has two different types of spin liquids in the two cases.

Most importantly, this has profound consequences on the anyon dynamics in the chiral phase of the model. The anyons in the AFM Kitaev model effectively move in the presence of a $\pi$ flux per unit cell \cite{prxvison,inti}. Furthermore, we have shown in Ref.~\cite{prxvison}, that for the gapless model ($\kappa=0$)
the vison hopping rate linear in $\Gamma$ or magnetic field vanishes for $K<0$ due to an interference effect. We will argue below, that a similar effect strongly suppresses the Raman signal.

For small perturbations, the Raman scattering amplitude, Eq.~\eqref{eq:raman_me},  is proportional to the matrix element 
\begin{align}
M^{h/\Gamma}_{n}\propto \bra{\Psi_0\left(0,\vec \delta^{(1/2)}_j\right)}\mathcal R^{h/\Gamma}\left({\vec \delta^{(1/2)}_j}\right)\ket{\Psi_0}\label{ramanM}
\end{align}
where $\ket{\Psi_0}$ is the groundstate of the pure Kitaev model and $\ket{\Psi_0\left(0,\vec \delta^{(1/2)}_j\right)}$ is the lowest-energy state of the pure Kitaev model with two anyons at position $0$ and $\vec \delta^{(1/2)}$. It is now important to realize that the Raman operators linking those two states always include two terms. For example, on a $z$ link $R^{\Gamma}$ is proportional to 
$\sigma_i^x \sigma_j^y+\sigma_i^y \sigma_j^x$. The two terms create the same two-anyon states with the amplitudes $A_1$ and $A_2$ . The two amplitudes are not independent of each other, but can be mapped onto each other by two different symmetries (inversion on the link and rotation by $\pi$ parallel to the link followed by time-reversal ).
For the ferromagnetic Kitaev model, $A_1=A_2$, and thus there is constructive interference. For the AFM model, in contrast, we find $A_1=-A_2$, the interference is destructive and the Raman matrix element vanishes exactly within our approximation
\begin{align}
M^{h/\Gamma}_{n}=0 \qquad \text{for }\ K<0.\label{ramanAFM}
\end{align}
As a consequence, the Raman signature of the vison bands will be much smaller in the AFM model compared to the FM case.

It is an interesting open question whether this result, which have been derived for 
the Raman couplings $\mathcal R^h$ and $\mathcal R^\Gamma$
 generalizes to other symmetry allowed Raman operators that create pairs of anyons. 
 We have checked that the same conclusion holds for all terms which can be written as  linear combinations of $\mathcal E_n  \mathcal E_m \sigma_i^\alpha$ or $\mathcal E_n  \mathcal E_m \sigma_i^\alpha \sigma_j^\beta$, where $i$ and $j$ are nearest neighbors.  This analysis does, however, not take into account the possible frequency dependent terms, and moreover ignores, e.g., magnetic dipole transitions.
Future work will be required to identify the leading-order processes which contribute to Raman spectrocopy in AFM Kitaev matter.

\section{Discussion}

In this paper, we have developed a general theory of how mobile anyons affect the Raman response in generic chiral Kitaev spin liquids. The Raman response is sensitive not only to the band-structure of anyons, but also to their interactions and braiding statistics. Most directly, the braiding statistics affects the power-law onset of Raman response. In the simplest setting (single band-minimum, linear polarization) we obtain the same result as has been derived by Morampudi {\it et al.} \cite{morampudi} for neutron scattering signatures of generic anyons. Extending there result, we have shown that the exponents depend not only on the polarization geometry but also on 
the number of minima in the single-anyon band structure. Our results are expected to apply directly also to other spectroscopic probes and other types of fractional excitations.

Another important property of anyons in Kitaev models is that they tend to form bound states and anti-bound states. There are two main mechanisms for bound-state formation: first, even in the pure Kitaev model there is a short-range interaction which is oscillating in sign. Second, nearest-neighbor pairs of anyons can also move more easily together then as single particles \cite{Batista,prxvison,inti,chen2024}. Thus, there is also an extra contribution to the binding energy arising directly from the motion of anyon-pairs. Bound states lead to sharp peaks in the Raman response with a  polarization dependence, which can be used as a fingerprint of the quantum numbers characterizing the bound state.

It is interesting to compare the  Raman response of anyons to the Raman signatures which can be expected from other possible degrees of freedom. Within our model, one can also expect a Raman signal from the gapped Majorana modes, which has recently been studied by Koller {\it et al.} \cite{koller2025}. While we assumed that the Majorana gap is larger than the anyonic gap, the situation can also be reversed. Massive Majorana modes can be viewed as fermions with their band minima at the $K$ and $K'$ points of the Brillouin zone. We can therefore use Eq.~\eqref{eq:twoBandMin} with $\alpha \to 1$ and obtain for the onset of the signal
\begin{align}
I_{LR}^\text{m}\sim I_{RL}^m\sim (\Delta \omega), \qquad I_{LL}^\text{m}\sim I_{RR}^m\sim (\Delta \omega)^0
\end{align}
with possible logarithmic corrections arising from (marginal) Majorana-Majorana interactions. This formula is consistent with the results of Ref.~\cite{koller2025}.
Thus, Majorana fermions can be easily distinguished from anyons, where the onset of the signal is strongly suppressed in the $RL$ and $LR$ channels. This is a consequence of the different statistics of anyons and Majorana modes.

It is more difficult, to distinguish the anyonic signature from those of a two magnon continuum which arise from gapped magnon bands as the power law exponent $\alpha=1/8$ is difficult to distinguish from $\alpha=0$. Due to the finite experimental resolution and disorder effects, a precise determination of the exponents characterizing the onset of the Raman signals is very challenging. Here, the most direct root to identify two-magnon continua is a simultaneous measurement of the one-magnon response either with Raman spectroscopy (where only $q=0$ modes can be seen), or with neutron scattering. Another useful signature can be the presence of bound states, which naturally arise in perturbed Kitaev models as discussed above. While magnons can also form bound states with similar quantum numbers, we are not aware of a generic mechanism for strong magnon attraction in perturbed Kitaev models.

Raman scattering studies of candidate Kitaev materials has been a very active area of experimental research \cite{PhysRevLett.113.187201,anuja1,anuja_chiral,wulferding2019raman}. The results of these studies are however strongly debated as their interpretation in terms of fractionalized Majorana fermions and visons remains ambiguous. Several works have also observed bound state signatures along with a continuum which are therefore interpreted as single magnons, magnon-bound states and multi magnon continua. A remarkable recent study by  Sahasrabudhe  {\it et al.} \cite{anuja_chiral} investigated, at intermediate magnetic fields, the Raman response and its polarization dependence.
The authors found that the Raman response was much larger in the $LR$ and $RL$ channels compared to the $LL/RR$ channels. At the same time, $I_{RL}$ and $I_{LR}$ were completely different in amplitude.
On a qualitative level, both of these features are consistent with our predictions, see Fig.~\ref{fig:continuum}. However, in the experimentally observed field range, the energy of the observed magnetic Raman modes always increases with  increasing magnetic fields. In contrast, one expects theoretically that our anyonic quasiparticles show the opposite trend: the gap decreases with field. This follows not only from perturbative calculations \cite{prxvison, Batista,chen2024} and numerical results \cite{ciaran}, but is also unavoidable  if one assumes that there is a second-order transition to a topologically trivial high-field phase. At such a phase transition, the anyon gap is expected to close.

While the unambiguous identification of spin liquids remains a challenging problem,  spectroscopic experiments remain the most direct way to explore exotic excitations and their interactions.
Many such experiments can be understood as creating pairs of fractionalized excitations with short range potential interactions and long-range statistical interactions. Therefore, we hope that our results, which account for these effects, can be useful for a broad range of experiments on fractionalized phases of matter.

\section*{Acknowledgments}
We thank Anuja Sahasrabudhe, Hamoon Hedayat and P. van Loosdrecht for useful discussions on their experimental data. A.P.J is grateful to Chuan Chen for helpful discussions on matrix element calculations and Natasha Perkins for useful discussions. This work was supported by the Deutsche Forschungsgemeinschaft (DFG) through CRC1238 (Project No. 277146847, project C02 and C04) and by the Bonn-Cologne Graduate School of Physics and Astronomy (BCGS).

\appendix
\section{Anyon-pair hopping matrix elements}
\label{appA}
Calculation of the hopping matrix elements and the associated Berry phases proceeds via the same Pfaffian method described in Ref.~\cite{prxvison}. However, some subtle difference between an $nn$ and $nnn$ pair hopping matrix elements require a proper discussion.

Consider the hopping matrix element of a (bosonic) $nn$ anyon pair from a $z$ bond $\nn{ij}$ to its adjacent $y$ bond $\nn{kj}$ (corresponding to the magenta hexagon in Fig.~\ref{fig:lattice}a). The hopping is induced at linear order in the external magnetic field and the corresponding matrix element is given by
\begin{align}
	t_{\textcolor{magenta}{\hexagon}} = h\bra{\Psi_0(0,\delta^{(1)}_2)}(\sigma^x_i+\sigma^x_j)\ket{\Psi_0(0,\delta^{(1)}_1)}
\end{align}
Here, one needs to remember that the (physical) ground state of a bosonic $nn$ anyon pair, requires the lowest matter Majorana mode, $a_0(\delta^{(1)}_1)$, to be occupied, implying
\begin{align} \ket{\Psi_0(0,\delta^{(1)}_1)} = \chi^\dagger_{\nn{ij},z}\ket{0}a_0^\dagger(\delta^{(1)}_1)\ket{M_0},\end{align}
where we have defined a `bond fermion' $\chi_{\nn{ij},\alpha}=b^\alpha_i+i b^\alpha_j$. After performing a gauge transformation and then contracting the bond fermions, the matrix elements become
\begin{align}
	t_{\textcolor{magenta}{\hexagon}} = h\bra{M_0(0,\delta^{(1)}_2)}a_0(\delta^{(1)}_2)(-i+c_ic_j)a_0^\dagger(\delta^{(1)}_1)\ket{M_0(0,\delta^{(1)}_1)}
\end{align}
For  a $\Gamma$ induced $nnn$ hopping of an $nnn$ anyon-pair (corresponding to a given side of the blue hexagon in Fig.~\ref{fig:lattice}a), the matrix elements can be shown to be of the form,
\begin{align}
	t_{\textcolor{blue}{\hexagon}} = \Gamma\bra{M_0(0,\delta^{(2)}_2)}(-1-ic_ic_j)\ket{M_0(0,\delta^{(2)}_1)}
\end{align}
note that, for the $nnn$ pair, the ground state has no extra Majorana mode occupied.

In contrast, when one calculates the $nnn$ hopping of an $nn$ anyon-pair (corresponding to a given side of the red triangles in Fig.~\ref{fig:lattice}a), the matrix elements are of the form,
\begin{align}
	t_{\textcolor{red}{\triangleleft}} = \Gamma\bra{M_0(0,\delta^{(1)}_3)}a_0(\delta^{(1)}_3)(-1-ic_ic_j)a_0^\dagger(\delta^{(1)}_1)\ket{M_0(0,\delta^{(1)}_1)}
\end{align}
The green triangles in Fig.~\ref{fig:lattice}a, involves transitions between an $nn$ and $nnn$ anyon-pair. The corresponding matrix element can be written as
 \begin{align}
 	t_{\textcolor{green}{\triangleright}} = h\bra{M_0(0,\delta^{(2)}_1)}(c_i+c_j)a_0^\dagger(\delta^{(1)}_1)\ket{M_0(0,\delta^{(1)}_1)}.
 \end{align}
 
The matrix elements given above can be then evaluated numerically on a finite size system using a generalized Wicks theorem  \cite{ring2004nuclear} which leads to Pfaffian determinants, as described in Ref.~\cite{prxvison,inti,MIZUSAKI2012219}. The hopping amplitudes show negligible finite size effects for a system of linear size $L\ge30$ when $\kappa \apprge 0.05 \,K$.

\newpage

\clearpage
\bibliography{references}

\providecommand{\noopsort}[1]{}\providecommand{\singleletter}[1]{#1}%
\begin{thebibliography}{45}%
\makeatletter
\providecommand \@ifxundefined [1]{%
 \@ifx{#1\undefined}
}%
\providecommand \@ifnum [1]{%
 \ifnum #1\expandafter \@firstoftwo
 \else \expandafter \@secondoftwo
 \fi
}%
\providecommand \@ifx [1]{%
 \ifx #1\expandafter \@firstoftwo
 \else \expandafter \@secondoftwo
 \fi
}%
\providecommand \natexlab [1]{#1}%
\providecommand \enquote  [1]{``#1''}%
\providecommand \bibnamefont  [1]{#1}%
\providecommand \bibfnamefont [1]{#1}%
\providecommand \citenamefont [1]{#1}%
\providecommand \href@noop [0]{\@secondoftwo}%
\providecommand \href [0]{\begingroup \@sanitize@url \@href}%
\providecommand \@href[1]{\@@startlink{#1}\@@href}%
\providecommand \@@href[1]{\endgroup#1\@@endlink}%
\providecommand \@sanitize@url [0]{\catcode `\\12\catcode `\$12\catcode
  `\&12\catcode `\#12\catcode `\^12\catcode `\_12\catcode `\%12\relax}%
\providecommand \@@startlink[1]{}%
\providecommand \@@endlink[0]{}%
\providecommand \url  [0]{\begingroup\@sanitize@url \@url }%
\providecommand \@url [1]{\endgroup\@href {#1}{\urlprefix }}%
\providecommand \urlprefix  [0]{URL }%
\providecommand \Eprint [0]{\href }%
\providecommand \doibase [0]{https://doi.org/}%
\providecommand \selectlanguage [0]{\@gobble}%
\providecommand \bibinfo  [0]{\@secondoftwo}%
\providecommand \bibfield  [0]{\@secondoftwo}%
\providecommand \translation [1]{[#1]}%
\providecommand \BibitemOpen [0]{}%
\providecommand \bibitemStop [0]{}%
\providecommand \bibitemNoStop [0]{.\EOS\space}%
\providecommand \EOS [0]{\spacefactor3000\relax}%
\providecommand \BibitemShut  [1]{\csname bibitem#1\endcsname}%
\let\auto@bib@innerbib\@empty
\bibitem [{\citenamefont {Leinaas}\ and\ \citenamefont
  {Myrheim}(1977)}]{leinaas1977theory}%
  \BibitemOpen
  \bibfield  {author} {\bibinfo {author} {\bibfnamefont {J.}~\bibnamefont
  {Leinaas}}\ and\ \bibinfo {author} {\bibfnamefont {J.}~\bibnamefont
  {Myrheim}},\ }\bibfield  {title} {\bibinfo {title} {On the theory of
  identical particles},\ }\href@noop {} {\bibfield  {journal} {\bibinfo
  {journal} {Il nuovo cimento}\ }\textbf {\bibinfo {volume} {37}},\ \bibinfo
  {pages} {132} (\bibinfo {year} {1977})}\BibitemShut {NoStop}%
\bibitem [{\citenamefont {Kitaev}(2006)}]{Kitaev06}%
  \BibitemOpen
  \bibfield  {author} {\bibinfo {author} {\bibfnamefont {A.}~\bibnamefont
  {Kitaev}},\ }\bibfield  {title} {\bibinfo {title} {{Anyons in an exactly
  solved model and beyond}},\ }\href
  {https://doi.org/https://doi.org/10.1016/j.aop.2005.10.005} {\bibfield
  {journal} {\bibinfo  {journal} {Annals of Physics}\ }\textbf {\bibinfo
  {volume} {321}},\ \bibinfo {pages} {2 } (\bibinfo {year} {2006})}\BibitemShut
  {NoStop}%
\bibitem [{\citenamefont {Khaliullin}\ and\ \citenamefont
  {Jackeli}(2009)}]{jackeli}%
  \BibitemOpen
  \bibfield  {author} {\bibinfo {author} {\bibfnamefont {G.}~\bibnamefont
  {Khaliullin}}\ and\ \bibinfo {author} {\bibfnamefont {G.}~\bibnamefont
  {Jackeli}},\ }\bibfield  {title} {\bibinfo {title} {{Mott Insulators in the
  Strong Spin-Orbit Coupling Limit: From Heisenberg to a Quantum Compass and
  Kitaev Models}},\ }\href
  {https://link.aps.org/doi/10.1103/PhysRevLett.102.017205} {\bibfield
  {journal} {\bibinfo  {journal} {Physical Review Letters}\ }\textbf {\bibinfo
  {volume} {102}} (\bibinfo {year} {2009})}\BibitemShut {NoStop}%
\bibitem [{\citenamefont {Rau}\ \emph {et~al.}(2014)\citenamefont {Rau},
  \citenamefont {Lee},\ and\ \citenamefont {Kee}}]{gammaRau}%
  \BibitemOpen
  \bibfield  {author} {\bibinfo {author} {\bibfnamefont {J.~G.}\ \bibnamefont
  {Rau}}, \bibinfo {author} {\bibfnamefont {E.~K.-H.}\ \bibnamefont {Lee}},\
  and\ \bibinfo {author} {\bibfnamefont {H.-Y.}\ \bibnamefont {Kee}},\
  }\bibfield  {title} {\bibinfo {title} {{Generic Spin Model for the Honeycomb
  Iridates beyond the Kitaev Limit}},\ }\href
  {https://doi.org/10.1103/PhysRevLett.112.077204} {\bibfield  {journal}
  {\bibinfo  {journal} {Phys. Rev. Lett.}\ }\textbf {\bibinfo {volume} {112}},\
  \bibinfo {pages} {077204} (\bibinfo {year} {2014})}\BibitemShut {NoStop}%
\bibitem [{\citenamefont {Chaloupka}\ \emph {et~al.}(2010)\citenamefont
  {Chaloupka}, \citenamefont {Jackeli},\ and\ \citenamefont
  {Khaliullin}}]{chaloupka2010kitaev}%
  \BibitemOpen
  \bibfield  {author} {\bibinfo {author} {\bibfnamefont {J.}~\bibnamefont
  {Chaloupka}}, \bibinfo {author} {\bibfnamefont {G.}~\bibnamefont {Jackeli}},\
  and\ \bibinfo {author} {\bibfnamefont {G.}~\bibnamefont {Khaliullin}},\
  }\bibfield  {title} {\bibinfo {title} {{Kitaev-Heisenberg Model on a
  Honeycomb Lattice: Possible Exotic Phases in Iridium Oxides A$_2$IrO$_3$}},\
  }\href@noop {} {\bibfield  {journal} {\bibinfo  {journal} {Physical review
  letters}\ }\textbf {\bibinfo {volume} {105}},\ \bibinfo {pages} {027204}
  (\bibinfo {year} {2010})}\BibitemShut {NoStop}%
\bibitem [{\citenamefont {Trebst}(2017{\natexlab{a}})}]{trebstreview}%
  \BibitemOpen
  \bibfield  {author} {\bibinfo {author} {\bibfnamefont {S.}~\bibnamefont
  {Trebst}},\ }\href@noop {} {\bibinfo {title} {{Kitaev Materials}}} (\bibinfo
  {year} {2017}{\natexlab{a}}),\ \Eprint {https://arxiv.org/abs/1701.07056}
  {arXiv:1701.07056 [cond-mat.str-el]} \BibitemShut {NoStop}%
\bibitem [{\citenamefont {Jan{\v{s}}a}\ \emph {et~al.}(2018)\citenamefont
  {Jan{\v{s}}a}, \citenamefont {Zorko}, \citenamefont {Gomil{\v{s}}ek},
  \citenamefont {Pregelj}, \citenamefont {Kr{\"a}mer}, \citenamefont {Biner},
  \citenamefont {Biffin}, \citenamefont {R{\"u}egg},\ and\ \citenamefont
  {Klanj{\v{s}}ek}}]{janvsa2018observation}%
  \BibitemOpen
  \bibfield  {author} {\bibinfo {author} {\bibfnamefont {N.}~\bibnamefont
  {Jan{\v{s}}a}}, \bibinfo {author} {\bibfnamefont {A.}~\bibnamefont {Zorko}},
  \bibinfo {author} {\bibfnamefont {M.}~\bibnamefont {Gomil{\v{s}}ek}},
  \bibinfo {author} {\bibfnamefont {M.}~\bibnamefont {Pregelj}}, \bibinfo
  {author} {\bibfnamefont {K.~W.}\ \bibnamefont {Kr{\"a}mer}}, \bibinfo
  {author} {\bibfnamefont {D.}~\bibnamefont {Biner}}, \bibinfo {author}
  {\bibfnamefont {A.}~\bibnamefont {Biffin}}, \bibinfo {author} {\bibfnamefont
  {C.}~\bibnamefont {R{\"u}egg}},\ and\ \bibinfo {author} {\bibfnamefont
  {M.}~\bibnamefont {Klanj{\v{s}}ek}},\ }\bibfield  {title} {\bibinfo {title}
  {Observation of two types of fractional excitation in the {K}itaev honeycomb
  magnet},\ }\href@noop {} {\bibfield  {journal} {\bibinfo  {journal} {Nature
  physics}\ }\textbf {\bibinfo {volume} {14}},\ \bibinfo {pages} {786}
  (\bibinfo {year} {2018})}\BibitemShut {NoStop}%
\bibitem [{\citenamefont {Moretti~Sala}\ \emph {et~al.}(2020)\citenamefont
  {Moretti~Sala}, \citenamefont {Monaco}, \citenamefont {Hickey}, \citenamefont
  {Becker}, \citenamefont {Freund}, \citenamefont {Jesche}, \citenamefont
  {Gegenwart}, \citenamefont {Eschmann}, \citenamefont {Buessen}, \citenamefont
  {Trebst}, \citenamefont {van Loosdrecht}, \citenamefont {van~den Brink},
  \citenamefont {Grüninger},\ and\ \citenamefont {Revelli}}]{nairo3}%
  \BibitemOpen
  \bibfield  {author} {\bibinfo {author} {\bibfnamefont {M.}~\bibnamefont
  {Moretti~Sala}}, \bibinfo {author} {\bibfnamefont {G.}~\bibnamefont
  {Monaco}}, \bibinfo {author} {\bibfnamefont {C.}~\bibnamefont {Hickey}},
  \bibinfo {author} {\bibfnamefont {P.}~\bibnamefont {Becker}}, \bibinfo
  {author} {\bibfnamefont {F.}~\bibnamefont {Freund}}, \bibinfo {author}
  {\bibfnamefont {A.}~\bibnamefont {Jesche}}, \bibinfo {author} {\bibfnamefont
  {P.}~\bibnamefont {Gegenwart}}, \bibinfo {author} {\bibfnamefont
  {T.}~\bibnamefont {Eschmann}}, \bibinfo {author} {\bibfnamefont {F.~L.}\
  \bibnamefont {Buessen}}, \bibinfo {author} {\bibfnamefont {S.}~\bibnamefont
  {Trebst}}, \bibinfo {author} {\bibfnamefont {P.~H.~M.}\ \bibnamefont {van
  Loosdrecht}}, \bibinfo {author} {\bibfnamefont {J.}~\bibnamefont {van~den
  Brink}}, \bibinfo {author} {\bibfnamefont {M.}~\bibnamefont {Grüninger}},\
  and\ \bibinfo {author} {\bibfnamefont {A.}~\bibnamefont {Revelli}},\
  }\bibfield  {title} {\bibinfo {title} {{Fingerprints of Kitaev physics in the
  magnetic excitations of honeycomb iridates}},\ }\href
  {https://link.aps.org/doi/10.1103/PhysRevResearch.2.043094} {\bibfield
  {journal} {\bibinfo  {journal} {Physical Review Research}\ }\textbf {\bibinfo
  {volume} {2}} (\bibinfo {year} {2020})}\BibitemShut {NoStop}%
\bibitem [{\citenamefont {Banerjee}\ \emph {et~al.}(2016)\citenamefont
  {Banerjee}, \citenamefont {Bridges}, \citenamefont {Yan}, \citenamefont
  {Aczel}, \citenamefont {Li}, \citenamefont {Stone}, \citenamefont {Granroth},
  \citenamefont {Lumsden}, \citenamefont {Yiu}, \citenamefont {Knolle},
  \citenamefont {Bhattacharjee}, \citenamefont {Kovrizhin}, \citenamefont
  {Moessner}, \citenamefont {Tennant}, \citenamefont {Mandrus},\ and\
  \citenamefont {Nagler}}]{proximate}%
  \BibitemOpen
  \bibfield  {author} {\bibinfo {author} {\bibfnamefont {A.}~\bibnamefont
  {Banerjee}}, \bibinfo {author} {\bibfnamefont {C.~A.}\ \bibnamefont
  {Bridges}}, \bibinfo {author} {\bibfnamefont {J.-Q.}\ \bibnamefont {Yan}},
  \bibinfo {author} {\bibfnamefont {A.~A.}\ \bibnamefont {Aczel}}, \bibinfo
  {author} {\bibfnamefont {L.}~\bibnamefont {Li}}, \bibinfo {author}
  {\bibfnamefont {M.~B.}\ \bibnamefont {Stone}}, \bibinfo {author}
  {\bibfnamefont {G.~E.}\ \bibnamefont {Granroth}}, \bibinfo {author}
  {\bibfnamefont {M.~D.}\ \bibnamefont {Lumsden}}, \bibinfo {author}
  {\bibfnamefont {Y.}~\bibnamefont {Yiu}}, \bibinfo {author} {\bibfnamefont
  {J.}~\bibnamefont {Knolle}}, \bibinfo {author} {\bibfnamefont
  {S.}~\bibnamefont {Bhattacharjee}}, \bibinfo {author} {\bibfnamefont {D.~L.}\
  \bibnamefont {Kovrizhin}}, \bibinfo {author} {\bibfnamefont {R.}~\bibnamefont
  {Moessner}}, \bibinfo {author} {\bibfnamefont {D.~A.}\ \bibnamefont
  {Tennant}}, \bibinfo {author} {\bibfnamefont {D.~G.}\ \bibnamefont
  {Mandrus}},\ and\ \bibinfo {author} {\bibfnamefont {S.~E.}\ \bibnamefont
  {Nagler}},\ }\bibfield  {title} {\bibinfo {title} {{Proximate Kitaev quantum
  spin liquid behaviour in a honeycomb magnet}},\ }\href@noop {} {\bibfield
  {journal} {\bibinfo  {journal} {Nature Materials}\ }\textbf {\bibinfo
  {volume} {15}} (\bibinfo {year} {2016})}\BibitemShut {NoStop}%
\bibitem [{\citenamefont {Banerjee}\ \emph {et~al.}(2018)\citenamefont
  {Banerjee}, \citenamefont {Lampen-Kelley}, \citenamefont {Knolle},
  \citenamefont {Balz}, \citenamefont {Aczel}, \citenamefont {Winn},
  \citenamefont {Liu}, \citenamefont {Pajerowski}, \citenamefont {Yan},
  \citenamefont {Bridges}, \citenamefont {Savici}, \citenamefont {Chakoumakos},
  \citenamefont {Lumsden}, \citenamefont {Tennant}, \citenamefont {Moessner},
  \citenamefont {Mandrus},\ and\ \citenamefont {Nagler}}]{fieldinduced}%
  \BibitemOpen
  \bibfield  {author} {\bibinfo {author} {\bibfnamefont {A.}~\bibnamefont
  {Banerjee}}, \bibinfo {author} {\bibfnamefont {P.}~\bibnamefont
  {Lampen-Kelley}}, \bibinfo {author} {\bibfnamefont {J.}~\bibnamefont
  {Knolle}}, \bibinfo {author} {\bibfnamefont {C.}~\bibnamefont {Balz}},
  \bibinfo {author} {\bibfnamefont {A.~A.}\ \bibnamefont {Aczel}}, \bibinfo
  {author} {\bibfnamefont {B.}~\bibnamefont {Winn}}, \bibinfo {author}
  {\bibfnamefont {Y.}~\bibnamefont {Liu}}, \bibinfo {author} {\bibfnamefont
  {D.}~\bibnamefont {Pajerowski}}, \bibinfo {author} {\bibfnamefont
  {J.}~\bibnamefont {Yan}}, \bibinfo {author} {\bibfnamefont {C.~A.}\
  \bibnamefont {Bridges}}, \bibinfo {author} {\bibfnamefont {A.~T.}\
  \bibnamefont {Savici}}, \bibinfo {author} {\bibfnamefont {B.~C.}\
  \bibnamefont {Chakoumakos}}, \bibinfo {author} {\bibfnamefont {M.~D.}\
  \bibnamefont {Lumsden}}, \bibinfo {author} {\bibfnamefont {D.~A.}\
  \bibnamefont {Tennant}}, \bibinfo {author} {\bibfnamefont {R.}~\bibnamefont
  {Moessner}}, \bibinfo {author} {\bibfnamefont {D.~G.}\ \bibnamefont
  {Mandrus}},\ and\ \bibinfo {author} {\bibfnamefont {S.~E.}\ \bibnamefont
  {Nagler}},\ }\bibfield  {title} {\bibinfo {title} {{Excitations in the
  field-induced quantum spin liquid state of $\alpha$-RuCl$_3$}},\ }\href
  {https://doi.org/10.1038/s41535-018-0079-2} {\bibfield  {journal} {\bibinfo
  {journal} {npj Quantum Materials}\ }\textbf {\bibinfo {volume} {3}},\
  \bibinfo {pages} {8} (\bibinfo {year} {2018})}\BibitemShut {NoStop}%
\bibitem [{\citenamefont {Kasahara}\ \emph {et~al.}(2018)\citenamefont
  {Kasahara}, \citenamefont {Ohnishi}, \citenamefont {Mizukami}, \citenamefont
  {Tanaka}, \citenamefont {Ma}, \citenamefont {Sugii}, \citenamefont {Kurita},
  \citenamefont {Tanaka}, \citenamefont {Nasu}, \citenamefont {Motome},
  \citenamefont {Shibauchi},\ and\ \citenamefont {Matsuda}}]{kasahara1}%
  \BibitemOpen
  \bibfield  {author} {\bibinfo {author} {\bibfnamefont {Y.}~\bibnamefont
  {Kasahara}}, \bibinfo {author} {\bibfnamefont {T.}~\bibnamefont {Ohnishi}},
  \bibinfo {author} {\bibfnamefont {Y.}~\bibnamefont {Mizukami}}, \bibinfo
  {author} {\bibfnamefont {O.}~\bibnamefont {Tanaka}}, \bibinfo {author}
  {\bibfnamefont {S.}~\bibnamefont {Ma}}, \bibinfo {author} {\bibfnamefont
  {K.}~\bibnamefont {Sugii}}, \bibinfo {author} {\bibfnamefont
  {N.}~\bibnamefont {Kurita}}, \bibinfo {author} {\bibfnamefont
  {H.}~\bibnamefont {Tanaka}}, \bibinfo {author} {\bibfnamefont
  {J.}~\bibnamefont {Nasu}}, \bibinfo {author} {\bibfnamefont {Y.}~\bibnamefont
  {Motome}}, \bibinfo {author} {\bibfnamefont {T.}~\bibnamefont {Shibauchi}},\
  and\ \bibinfo {author} {\bibfnamefont {Y.}~\bibnamefont {Matsuda}},\
  }\bibfield  {title} {\bibinfo {title} {{Majorana quantization and
  half-integer thermal quantum Hall effect in a Kitaev spin liquid}},\
  }\href@noop {} {\bibfield  {journal} {\bibinfo  {journal} {Nature}\ }\textbf
  {\bibinfo {volume} {559}} (\bibinfo {year} {2018})}\BibitemShut {NoStop}%
\bibitem [{\citenamefont {Yokoi}\ \emph {et~al.}(2021)\citenamefont {Yokoi},
  \citenamefont {Ma}, \citenamefont {Kasahara}, \citenamefont {Kasahara},
  \citenamefont {Shibauchi}, \citenamefont {Kurita}, \citenamefont {Tanaka},
  \citenamefont {Nasu}, \citenamefont {Motome}, \citenamefont {Hickey},
  \citenamefont {Trebst},\ and\ \citenamefont {Matsuda}}]{kasahara2}%
  \BibitemOpen
  \bibfield  {author} {\bibinfo {author} {\bibfnamefont {T.}~\bibnamefont
  {Yokoi}}, \bibinfo {author} {\bibfnamefont {S.}~\bibnamefont {Ma}}, \bibinfo
  {author} {\bibfnamefont {Y.}~\bibnamefont {Kasahara}}, \bibinfo {author}
  {\bibfnamefont {S.}~\bibnamefont {Kasahara}}, \bibinfo {author}
  {\bibfnamefont {T.}~\bibnamefont {Shibauchi}}, \bibinfo {author}
  {\bibfnamefont {N.}~\bibnamefont {Kurita}}, \bibinfo {author} {\bibfnamefont
  {H.}~\bibnamefont {Tanaka}}, \bibinfo {author} {\bibfnamefont
  {J.}~\bibnamefont {Nasu}}, \bibinfo {author} {\bibfnamefont {Y.}~\bibnamefont
  {Motome}}, \bibinfo {author} {\bibfnamefont {C.}~\bibnamefont {Hickey}},
  \bibinfo {author} {\bibfnamefont {S.}~\bibnamefont {Trebst}},\ and\ \bibinfo
  {author} {\bibfnamefont {Y.}~\bibnamefont {Matsuda}},\ }\bibfield  {title}
  {\bibinfo {title} {{Half-integer quantized anomalous thermal Hall effect in
  the Kitaev material candidate $\alpha$-RuCl$_3$}},\ }\href
  {http://science.sciencemag.org/content/373/6554/568.abstract} {\bibfield
  {journal} {\bibinfo  {journal} {Science}\ }\textbf {\bibinfo {volume} {373}}
  (\bibinfo {year} {2021})}\BibitemShut {NoStop}%
\bibitem [{\citenamefont {Morampudi}\ \emph {et~al.}(2017)\citenamefont
  {Morampudi}, \citenamefont {Turner}, \citenamefont {Pollmann},\ and\
  \citenamefont {Wilczek}}]{morampudi}%
  \BibitemOpen
  \bibfield  {author} {\bibinfo {author} {\bibfnamefont {S.~C.}\ \bibnamefont
  {Morampudi}}, \bibinfo {author} {\bibfnamefont {A.~M.}\ \bibnamefont
  {Turner}}, \bibinfo {author} {\bibfnamefont {F.}~\bibnamefont {Pollmann}},\
  and\ \bibinfo {author} {\bibfnamefont {F.}~\bibnamefont {Wilczek}},\
  }\bibfield  {title} {\bibinfo {title} {Statistics of fractionalized
  excitations through threshold spectroscopy},\ }\href
  {https://doi.org/10.1103/PhysRevLett.118.227201} {\bibfield  {journal}
  {\bibinfo  {journal} {Phys. Rev. Lett.}\ }\textbf {\bibinfo {volume} {118}},\
  \bibinfo {pages} {227201} (\bibinfo {year} {2017})}\BibitemShut {NoStop}%
\bibitem [{\citenamefont {Wulferding}\ \emph {et~al.}(2019)\citenamefont
  {Wulferding}, \citenamefont {Choi}, \citenamefont {Lee},\ and\ \citenamefont
  {Choi}}]{wulferding2019raman}%
  \BibitemOpen
  \bibfield  {author} {\bibinfo {author} {\bibfnamefont {D.}~\bibnamefont
  {Wulferding}}, \bibinfo {author} {\bibfnamefont {Y.}~\bibnamefont {Choi}},
  \bibinfo {author} {\bibfnamefont {W.}~\bibnamefont {Lee}},\ and\ \bibinfo
  {author} {\bibfnamefont {K.-Y.}\ \bibnamefont {Choi}},\ }\bibfield  {title}
  {\bibinfo {title} {Raman spectroscopic diagnostic of quantum spin liquids},\
  }\href@noop {} {\bibfield  {journal} {\bibinfo  {journal} {Journal of
  Physics: Condensed Matter}\ }\textbf {\bibinfo {volume} {32}},\ \bibinfo
  {pages} {043001} (\bibinfo {year} {2019})}\BibitemShut {NoStop}%
\bibitem [{\citenamefont {Choi}\ \emph {et~al.}(2021)\citenamefont {Choi},
  \citenamefont {Lee}, \citenamefont {Lee}, \citenamefont {Lee}, \citenamefont
  {Seong},\ and\ \citenamefont {Choi}}]{choi2021bosonic}%
  \BibitemOpen
  \bibfield  {author} {\bibinfo {author} {\bibfnamefont {Y.}~\bibnamefont
  {Choi}}, \bibinfo {author} {\bibfnamefont {S.}~\bibnamefont {Lee}}, \bibinfo
  {author} {\bibfnamefont {J.-H.}\ \bibnamefont {Lee}}, \bibinfo {author}
  {\bibfnamefont {S.}~\bibnamefont {Lee}}, \bibinfo {author} {\bibfnamefont
  {M.-J.}\ \bibnamefont {Seong}},\ and\ \bibinfo {author} {\bibfnamefont
  {K.-Y.}\ \bibnamefont {Choi}},\ }\bibfield  {title} {\bibinfo {title}
  {Bosonic spinons in anisotropic triangular antiferromagnets},\ }\href@noop {}
  {\bibfield  {journal} {\bibinfo  {journal} {Nature Communications}\ }\textbf
  {\bibinfo {volume} {12}},\ \bibinfo {pages} {6453} (\bibinfo {year}
  {2021})}\BibitemShut {NoStop}%
\bibitem [{\citenamefont {Misochko}\ \emph {et~al.}(1996)\citenamefont
  {Misochko}, \citenamefont {Tajima}, \citenamefont {Urano}, \citenamefont
  {Eisaki},\ and\ \citenamefont {Uchida}}]{PhysRevB.53.R14733}%
  \BibitemOpen
  \bibfield  {author} {\bibinfo {author} {\bibfnamefont {O.~V.}\ \bibnamefont
  {Misochko}}, \bibinfo {author} {\bibfnamefont {S.}~\bibnamefont {Tajima}},
  \bibinfo {author} {\bibfnamefont {C.}~\bibnamefont {Urano}}, \bibinfo
  {author} {\bibfnamefont {H.}~\bibnamefont {Eisaki}},\ and\ \bibinfo {author}
  {\bibfnamefont {S.}~\bibnamefont {Uchida}},\ }\bibfield  {title} {\bibinfo
  {title} {{Raman-scattering evidence for free spinons in the one-dimensional
  spin-\textonehalf{} chains of ${\mathrm{Sr}}_{2}$Cu${\mathrm{O}}_{3}$ and
  SrCu${\mathrm{O}}_{2}$}},\ }\href
  {https://doi.org/10.1103/PhysRevB.53.R14733} {\bibfield  {journal} {\bibinfo
  {journal} {Phys. Rev. B}\ }\textbf {\bibinfo {volume} {53}},\ \bibinfo
  {pages} {R14733} (\bibinfo {year} {1996})}\BibitemShut {NoStop}%
\bibitem [{\citenamefont {Gnezdilov}\ \emph {et~al.}(2012)\citenamefont
  {Gnezdilov}, \citenamefont {Deisenhofer}, \citenamefont {Lemmens},
  \citenamefont {Wulferding}, \citenamefont {Afanasiev}, \citenamefont
  {Ghigna}, \citenamefont {Loidl},\ and\ \citenamefont
  {Yeremenko}}]{gnezdilov2012phononic}%
  \BibitemOpen
  \bibfield  {author} {\bibinfo {author} {\bibfnamefont {V.}~\bibnamefont
  {Gnezdilov}}, \bibinfo {author} {\bibfnamefont {J.}~\bibnamefont
  {Deisenhofer}}, \bibinfo {author} {\bibfnamefont {P.}~\bibnamefont
  {Lemmens}}, \bibinfo {author} {\bibfnamefont {D.}~\bibnamefont {Wulferding}},
  \bibinfo {author} {\bibfnamefont {O.}~\bibnamefont {Afanasiev}}, \bibinfo
  {author} {\bibfnamefont {P.}~\bibnamefont {Ghigna}}, \bibinfo {author}
  {\bibfnamefont {A.}~\bibnamefont {Loidl}},\ and\ \bibinfo {author}
  {\bibfnamefont {A.}~\bibnamefont {Yeremenko}},\ }\bibfield  {title} {\bibinfo
  {title} {{Phononic and magnetic excitations in the quasi-one-dimensional
  Heisenberg antiferromagnet KCuF$_3$}},\ }\href@noop {} {\bibfield  {journal}
  {\bibinfo  {journal} {Low Temperature Physics}\ }\textbf {\bibinfo {volume}
  {38}},\ \bibinfo {pages} {419} (\bibinfo {year} {2012})}\BibitemShut
  {NoStop}%
\bibitem [{\citenamefont {Ko}\ \emph {et~al.}(2010)\citenamefont {Ko},
  \citenamefont {Liu}, \citenamefont {Ng},\ and\ \citenamefont
  {Lee}}]{raman_u1}%
  \BibitemOpen
  \bibfield  {author} {\bibinfo {author} {\bibfnamefont {W.-H.}\ \bibnamefont
  {Ko}}, \bibinfo {author} {\bibfnamefont {Z.-X.}\ \bibnamefont {Liu}},
  \bibinfo {author} {\bibfnamefont {T.-K.}\ \bibnamefont {Ng}},\ and\ \bibinfo
  {author} {\bibfnamefont {P.~A.}\ \bibnamefont {Lee}},\ }\bibfield  {title}
  {\bibinfo {title} {{Raman signature of the U(1) {D}irac spin-liquid state in
  the spin-$\frac{1}{2}$ kagome system}},\ }\href
  {https://doi.org/10.1103/PhysRevB.81.024414} {\bibfield  {journal} {\bibinfo
  {journal} {Phys. Rev. B}\ }\textbf {\bibinfo {volume} {81}},\ \bibinfo
  {pages} {024414} (\bibinfo {year} {2010})}\BibitemShut {NoStop}%
\bibitem [{\citenamefont {Tang}\ \emph {et~al.}(2022)\citenamefont {Tang},
  \citenamefont {Moritz},\ and\ \citenamefont
  {Devereaux}}]{PhysRevB.106.064428}%
  \BibitemOpen
  \bibfield  {author} {\bibinfo {author} {\bibfnamefont {T.}~\bibnamefont
  {Tang}}, \bibinfo {author} {\bibfnamefont {B.}~\bibnamefont {Moritz}},\ and\
  \bibinfo {author} {\bibfnamefont {T.~P.}\ \bibnamefont {Devereaux}},\
  }\bibfield  {title} {\bibinfo {title} {Spectra of a gapped quantum spin
  liquid with a strong chiral excitation on the triangular lattice},\ }\href
  {https://doi.org/10.1103/PhysRevB.106.064428} {\bibfield  {journal} {\bibinfo
   {journal} {Phys. Rev. B}\ }\textbf {\bibinfo {volume} {106}},\ \bibinfo
  {pages} {064428} (\bibinfo {year} {2022})}\BibitemShut {NoStop}%
\bibitem [{\citenamefont {Knolle}\ \emph {et~al.}(2014)\citenamefont {Knolle},
  \citenamefont {Chern}, \citenamefont {Kovrizhin}, \citenamefont {Moessner},\
  and\ \citenamefont {Perkins}}]{PhysRevLett.113.187201}%
  \BibitemOpen
  \bibfield  {author} {\bibinfo {author} {\bibfnamefont {J.}~\bibnamefont
  {Knolle}}, \bibinfo {author} {\bibfnamefont {G.-W.}\ \bibnamefont {Chern}},
  \bibinfo {author} {\bibfnamefont {D.~L.}\ \bibnamefont {Kovrizhin}}, \bibinfo
  {author} {\bibfnamefont {R.}~\bibnamefont {Moessner}},\ and\ \bibinfo
  {author} {\bibfnamefont {N.~B.}\ \bibnamefont {Perkins}},\ }\bibfield
  {title} {\bibinfo {title} {{Raman Scattering Signatures of {K}itaev Spin
  Liquids in ${A}_{2}{\mathrm{IrO}}_{3}$ {I}ridates with $A=\mathrm{Na}$ or
  Li}},\ }\href {https://doi.org/10.1103/PhysRevLett.113.187201} {\bibfield
  {journal} {\bibinfo  {journal} {Phys. Rev. Lett.}\ }\textbf {\bibinfo
  {volume} {113}},\ \bibinfo {pages} {187201} (\bibinfo {year}
  {2014})}\BibitemShut {NoStop}%
\bibitem [{\citenamefont {Sandilands}\ \emph {et~al.}(2015)\citenamefont
  {Sandilands}, \citenamefont {Tian}, \citenamefont {Plumb}, \citenamefont
  {Kim},\ and\ \citenamefont {Burch}}]{sandilands2015scattering}%
  \BibitemOpen
  \bibfield  {author} {\bibinfo {author} {\bibfnamefont {L.~J.}\ \bibnamefont
  {Sandilands}}, \bibinfo {author} {\bibfnamefont {Y.}~\bibnamefont {Tian}},
  \bibinfo {author} {\bibfnamefont {K.~W.}\ \bibnamefont {Plumb}}, \bibinfo
  {author} {\bibfnamefont {Y.-J.}\ \bibnamefont {Kim}},\ and\ \bibinfo {author}
  {\bibfnamefont {K.~S.}\ \bibnamefont {Burch}},\ }\bibfield  {title} {\bibinfo
  {title} {{Scattering continuum and possible fractionalized excitations in
  $\alpha$-RuCl$_3$}},\ }\href@noop {} {\bibfield  {journal} {\bibinfo
  {journal} {Physical review letters}\ }\textbf {\bibinfo {volume} {114}},\
  \bibinfo {pages} {147201} (\bibinfo {year} {2015})}\BibitemShut {NoStop}%
\bibitem [{\citenamefont {Nasu}\ \emph {et~al.}(2016)\citenamefont {Nasu},
  \citenamefont {Knolle}, \citenamefont {Kovrizhin}, \citenamefont {Motome},\
  and\ \citenamefont {Moessner}}]{nasu2016fermionic}%
  \BibitemOpen
  \bibfield  {author} {\bibinfo {author} {\bibfnamefont {J.}~\bibnamefont
  {Nasu}}, \bibinfo {author} {\bibfnamefont {J.}~\bibnamefont {Knolle}},
  \bibinfo {author} {\bibfnamefont {D.~L.}\ \bibnamefont {Kovrizhin}}, \bibinfo
  {author} {\bibfnamefont {Y.}~\bibnamefont {Motome}},\ and\ \bibinfo {author}
  {\bibfnamefont {R.}~\bibnamefont {Moessner}},\ }\bibfield  {title} {\bibinfo
  {title} {Fermionic response from fractionalization in an insulating
  two-dimensional magnet},\ }\href@noop {} {\bibfield  {journal} {\bibinfo
  {journal} {Nature Physics}\ }\textbf {\bibinfo {volume} {12}},\ \bibinfo
  {pages} {912} (\bibinfo {year} {2016})}\BibitemShut {NoStop}%
\bibitem [{\citenamefont {Wulferding}\ \emph {et~al.}(2020)\citenamefont
  {Wulferding}, \citenamefont {Choi}, \citenamefont {Do}, \citenamefont {Lee},
  \citenamefont {Lemmens}, \citenamefont {Faugeras}, \citenamefont {Gallais},\
  and\ \citenamefont {Choi}}]{wulferding2020magnon}%
  \BibitemOpen
  \bibfield  {author} {\bibinfo {author} {\bibfnamefont {D.}~\bibnamefont
  {Wulferding}}, \bibinfo {author} {\bibfnamefont {Y.}~\bibnamefont {Choi}},
  \bibinfo {author} {\bibfnamefont {S.-H.}\ \bibnamefont {Do}}, \bibinfo
  {author} {\bibfnamefont {C.~H.}\ \bibnamefont {Lee}}, \bibinfo {author}
  {\bibfnamefont {P.}~\bibnamefont {Lemmens}}, \bibinfo {author} {\bibfnamefont
  {C.}~\bibnamefont {Faugeras}}, \bibinfo {author} {\bibfnamefont
  {Y.}~\bibnamefont {Gallais}},\ and\ \bibinfo {author} {\bibfnamefont {K.-Y.}\
  \bibnamefont {Choi}},\ }\bibfield  {title} {\bibinfo {title} {Magnon bound
  states versus anyonic {M}ajorana excitations in the {K}itaev honeycomb magnet
  $\alpha$-{R}u{C}l$_3$},\ }\href@noop {} {\bibfield  {journal} {\bibinfo
  {journal} {Nature communications}\ }\textbf {\bibinfo {volume} {11}},\
  \bibinfo {pages} {1603} (\bibinfo {year} {2020})}\BibitemShut {NoStop}%
\bibitem [{\citenamefont {Sahasrabudhe}\ \emph {et~al.}(2020)\citenamefont
  {Sahasrabudhe}, \citenamefont {Kaib}, \citenamefont {Reschke}, \citenamefont
  {German}, \citenamefont {Koethe}, \citenamefont {Buhot}, \citenamefont
  {Kamenskyi}, \citenamefont {Hickey}, \citenamefont {Becker}, \citenamefont
  {Tsurkan}, \citenamefont {Loidl}, \citenamefont {Do}, \citenamefont {Choi},
  \citenamefont {Gr\"uninger}, \citenamefont {Winter}, \citenamefont {Wang},
  \citenamefont {Valent\'{\i}},\ and\ \citenamefont {van Loosdrecht}}]{anuja1}%
  \BibitemOpen
  \bibfield  {author} {\bibinfo {author} {\bibfnamefont {A.}~\bibnamefont
  {Sahasrabudhe}}, \bibinfo {author} {\bibfnamefont {D.~A.~S.}\ \bibnamefont
  {Kaib}}, \bibinfo {author} {\bibfnamefont {S.}~\bibnamefont {Reschke}},
  \bibinfo {author} {\bibfnamefont {R.}~\bibnamefont {German}}, \bibinfo
  {author} {\bibfnamefont {T.~C.}\ \bibnamefont {Koethe}}, \bibinfo {author}
  {\bibfnamefont {J.}~\bibnamefont {Buhot}}, \bibinfo {author} {\bibfnamefont
  {D.}~\bibnamefont {Kamenskyi}}, \bibinfo {author} {\bibfnamefont
  {C.}~\bibnamefont {Hickey}}, \bibinfo {author} {\bibfnamefont
  {P.}~\bibnamefont {Becker}}, \bibinfo {author} {\bibfnamefont
  {V.}~\bibnamefont {Tsurkan}}, \bibinfo {author} {\bibfnamefont
  {A.}~\bibnamefont {Loidl}}, \bibinfo {author} {\bibfnamefont {S.~H.}\
  \bibnamefont {Do}}, \bibinfo {author} {\bibfnamefont {K.~Y.}\ \bibnamefont
  {Choi}}, \bibinfo {author} {\bibfnamefont {M.}~\bibnamefont {Gr\"uninger}},
  \bibinfo {author} {\bibfnamefont {S.~M.}\ \bibnamefont {Winter}}, \bibinfo
  {author} {\bibfnamefont {Z.}~\bibnamefont {Wang}}, \bibinfo {author}
  {\bibfnamefont {R.}~\bibnamefont {Valent\'{\i}}},\ and\ \bibinfo {author}
  {\bibfnamefont {P.~H.~M.}\ \bibnamefont {van Loosdrecht}},\ }\bibfield
  {title} {\bibinfo {title} {{High-field quantum disordered state in
  $\ensuremath{\alpha}\ensuremath{-}{\mathrm{RuCl}}_{3}$: Spin flips, bound
  states, and multiparticle continuum}},\ }\href
  {https://doi.org/10.1103/PhysRevB.101.140410} {\bibfield  {journal} {\bibinfo
   {journal} {Phys. Rev. B}\ }\textbf {\bibinfo {volume} {101}},\ \bibinfo
  {pages} {140410} (\bibinfo {year} {2020})}\BibitemShut {NoStop}%
\bibitem [{\citenamefont {Sahasrabudhe}\ \emph {et~al.}(2024)\citenamefont
  {Sahasrabudhe}, \citenamefont {Prosnikov}, \citenamefont {Koethe},
  \citenamefont {Stein}, \citenamefont {Tsurkan}, \citenamefont {Loidl},
  \citenamefont {Gr\"uninger}, \citenamefont {Hedayat},\ and\ \citenamefont
  {van Loosdrecht}}]{anuja_chiral}%
  \BibitemOpen
  \bibfield  {author} {\bibinfo {author} {\bibfnamefont {A.}~\bibnamefont
  {Sahasrabudhe}}, \bibinfo {author} {\bibfnamefont {M.~A.}\ \bibnamefont
  {Prosnikov}}, \bibinfo {author} {\bibfnamefont {T.~C.}\ \bibnamefont
  {Koethe}}, \bibinfo {author} {\bibfnamefont {P.}~\bibnamefont {Stein}},
  \bibinfo {author} {\bibfnamefont {V.}~\bibnamefont {Tsurkan}}, \bibinfo
  {author} {\bibfnamefont {A.}~\bibnamefont {Loidl}}, \bibinfo {author}
  {\bibfnamefont {M.}~\bibnamefont {Gr\"uninger}}, \bibinfo {author}
  {\bibfnamefont {H.}~\bibnamefont {Hedayat}},\ and\ \bibinfo {author}
  {\bibfnamefont {P.~H.~M.}\ \bibnamefont {van Loosdrecht}},\ }\bibfield
  {title} {\bibinfo {title} {{Chiral excitations and the intermediate-field
  regime in the {K}itaev magnet
  $\ensuremath{\alpha}\ensuremath{-}{\mathrm{RuCl}}_{3}$}},\ }\href
  {https://doi.org/10.1103/PhysRevResearch.6.L022005} {\bibfield  {journal}
  {\bibinfo  {journal} {Phys. Rev. Res.}\ }\textbf {\bibinfo {volume} {6}},\
  \bibinfo {pages} {L022005} (\bibinfo {year} {2024})}\BibitemShut {NoStop}%
\bibitem [{\citenamefont {Fleury}\ and\ \citenamefont
  {Loudon}(1968)}]{Loudon_Fleury}%
  \BibitemOpen
  \bibfield  {author} {\bibinfo {author} {\bibfnamefont {P.~A.}\ \bibnamefont
  {Fleury}}\ and\ \bibinfo {author} {\bibfnamefont {R.}~\bibnamefont
  {Loudon}},\ }\bibfield  {title} {\bibinfo {title} {Scattering of light by
  one- and two-magnon excitations},\ }\href
  {https://doi.org/10.1103/PhysRev.166.514} {\bibfield  {journal} {\bibinfo
  {journal} {Phys. Rev.}\ }\textbf {\bibinfo {volume} {166}},\ \bibinfo {pages}
  {514} (\bibinfo {year} {1968})}\BibitemShut {NoStop}%
\bibitem [{\citenamefont {Yang}\ \emph {et~al.}(2021)\citenamefont {Yang},
  \citenamefont {Li}, \citenamefont {Rousochatzakis},\ and\ \citenamefont
  {Perkins}}]{nonLF_perkins}%
  \BibitemOpen
  \bibfield  {author} {\bibinfo {author} {\bibfnamefont {Y.}~\bibnamefont
  {Yang}}, \bibinfo {author} {\bibfnamefont {M.}~\bibnamefont {Li}}, \bibinfo
  {author} {\bibfnamefont {I.}~\bibnamefont {Rousochatzakis}},\ and\ \bibinfo
  {author} {\bibfnamefont {N.~B.}\ \bibnamefont {Perkins}},\ }\bibfield
  {title} {\bibinfo {title} {Non-loudon-fleury {R}aman scattering in spin-orbit
  coupled {M}ott insulators},\ }\href
  {https://doi.org/10.1103/PhysRevB.104.144412} {\bibfield  {journal} {\bibinfo
   {journal} {Phys. Rev. B}\ }\textbf {\bibinfo {volume} {104}},\ \bibinfo
  {pages} {144412} (\bibinfo {year} {2021})}\BibitemShut {NoStop}%
\bibitem [{\citenamefont {Lahtinen}\ \emph {et~al.}(2012)\citenamefont
  {Lahtinen}, \citenamefont {Ludwig}, \citenamefont {Pachos},\ and\
  \citenamefont {Trebst}}]{topoliquid}%
  \BibitemOpen
  \bibfield  {author} {\bibinfo {author} {\bibfnamefont {V.}~\bibnamefont
  {Lahtinen}}, \bibinfo {author} {\bibfnamefont {A.~W.~W.}\ \bibnamefont
  {Ludwig}}, \bibinfo {author} {\bibfnamefont {J.~K.}\ \bibnamefont {Pachos}},\
  and\ \bibinfo {author} {\bibfnamefont {S.}~\bibnamefont {Trebst}},\
  }\bibfield  {title} {\bibinfo {title} {Topological liquid nucleation induced
  by vortex-vortex interactions in kitaev's honeycomb model},\ }\href
  {https://doi.org/10.1103/PhysRevB.86.075115} {\bibfield  {journal} {\bibinfo
  {journal} {Phys. Rev. B}\ }\textbf {\bibinfo {volume} {86}},\ \bibinfo
  {pages} {075115} (\bibinfo {year} {2012})}\BibitemShut {NoStop}%
\bibitem [{\citenamefont {Lahtinen}(2011)}]{lahtinennjp}%
  \BibitemOpen
  \bibfield  {author} {\bibinfo {author} {\bibfnamefont {V.}~\bibnamefont
  {Lahtinen}},\ }\bibfield  {title} {\bibinfo {title} {Interacting non-abelian
  anyons as majorana fermions in the honeycomb lattice model},\ }\href
  {https://doi.org/10.1088/1367-2630/13/7/075009} {\bibfield  {journal}
  {\bibinfo  {journal} {New Journal of Physics}\ }\textbf {\bibinfo {volume}
  {13}},\ \bibinfo {pages} {075009} (\bibinfo {year} {2011})}\BibitemShut
  {NoStop}%
\bibitem [{\citenamefont {Pachos}(2012)}]{Pachos_2012}%
  \BibitemOpen
  \bibfield  {author} {\bibinfo {author} {\bibfnamefont {J.~K.}\ \bibnamefont
  {Pachos}},\ }\href@noop {} {\emph {\bibinfo {title} {Introduction to
  Topological Quantum Computation}}}\ (\bibinfo  {publisher} {Cambridge
  University Press},\ \bibinfo {year} {2012})\BibitemShut {NoStop}%
\bibitem [{\citenamefont {Knolle}(2016)}]{Knollethesis}%
  \BibitemOpen
  \bibfield  {author} {\bibinfo {author} {\bibfnamefont {J.}~\bibnamefont
  {Knolle}},\ }\bibfield  {title} {\bibinfo {title} {{Dynamics of a Quantum
  Spin Liquid}}\ }(\bibinfo {year} {2016})\BibitemShut {NoStop}%
\bibitem [{\citenamefont {Trebst}(2017{\natexlab{b}})}]{trebst2017kitaev}%
  \BibitemOpen
  \bibfield  {author} {\bibinfo {author} {\bibfnamefont {S.}~\bibnamefont
  {Trebst}},\ }\href@noop {} {\bibinfo {title} {{Kitaev {M}aterials}}}
  (\bibinfo {year} {2017}{\natexlab{b}}),\ \Eprint
  {https://arxiv.org/abs/1701.07056} {arXiv:1701.07056 [cond-mat.str-el]}
  \BibitemShut {NoStop}%
\bibitem [{\citenamefont {Joy}\ and\ \citenamefont {Rosch}(2022)}]{prxvison}%
  \BibitemOpen
  \bibfield  {author} {\bibinfo {author} {\bibfnamefont {A.~P.}\ \bibnamefont
  {Joy}}\ and\ \bibinfo {author} {\bibfnamefont {A.}~\bibnamefont {Rosch}},\
  }\bibfield  {title} {\bibinfo {title} {Dynamics of visons and thermal hall
  effect in perturbed kitaev models},\ }\href
  {https://doi.org/10.1103/PhysRevX.12.041004} {\bibfield  {journal} {\bibinfo
  {journal} {Phys. Rev. X}\ }\textbf {\bibinfo {volume} {12}},\ \bibinfo
  {pages} {041004} (\bibinfo {year} {2022})}\BibitemShut {NoStop}%
\bibitem [{\citenamefont {Zhang}\ \emph {et~al.}(2021)\citenamefont {Zhang},
  \citenamefont {Hal\'asz}, \citenamefont {Zhu},\ and\ \citenamefont
  {Batista}}]{Batista}%
  \BibitemOpen
  \bibfield  {author} {\bibinfo {author} {\bibfnamefont {S.-S.}\ \bibnamefont
  {Zhang}}, \bibinfo {author} {\bibfnamefont {G.~B.}\ \bibnamefont {Hal\'asz}},
  \bibinfo {author} {\bibfnamefont {W.}~\bibnamefont {Zhu}},\ and\ \bibinfo
  {author} {\bibfnamefont {C.~D.}\ \bibnamefont {Batista}},\ }\bibfield
  {title} {\bibinfo {title} {{Variational study of the Kitaev-Heisenberg-Gamma
  model}},\ }\href {https://doi.org/10.1103/PhysRevB.104.014411} {\bibfield
  {journal} {\bibinfo  {journal} {Phys. Rev. B}\ }\textbf {\bibinfo {volume}
  {104}},\ \bibinfo {pages} {014411} (\bibinfo {year} {2021})}\BibitemShut
  {NoStop}%
\bibitem [{\citenamefont {Joy}\ and\ \citenamefont
  {Rosch}(2024)}]{joy2024gauge}%
  \BibitemOpen
  \bibfield  {author} {\bibinfo {author} {\bibfnamefont {A.~P.}\ \bibnamefont
  {Joy}}\ and\ \bibinfo {author} {\bibfnamefont {A.}~\bibnamefont {Rosch}},\
  }\bibfield  {title} {\bibinfo {title} {Gauge field dynamics in multilayer
  kitaev spin liquids},\ }\href@noop {} {\bibfield  {journal} {\bibinfo
  {journal} {npj Quantum Materials}\ }\textbf {\bibinfo {volume} {9}},\
  \bibinfo {pages} {62} (\bibinfo {year} {2024})}\BibitemShut {NoStop}%
\bibitem [{\citenamefont {Pozo}\ \emph {et~al.}(2021)\citenamefont {Pozo},
  \citenamefont {Rao}, \citenamefont {Chen},\ and\ \citenamefont
  {Sodemann}}]{inti}%
  \BibitemOpen
  \bibfield  {author} {\bibinfo {author} {\bibfnamefont {O.}~\bibnamefont
  {Pozo}}, \bibinfo {author} {\bibfnamefont {P.}~\bibnamefont {Rao}}, \bibinfo
  {author} {\bibfnamefont {C.}~\bibnamefont {Chen}},\ and\ \bibinfo {author}
  {\bibfnamefont {I.}~\bibnamefont {Sodemann}},\ }\bibfield  {title} {\bibinfo
  {title} {{Anatomy of ${\mathbb{Z}}_{2}$ fluxes in anyon Fermi liquids and
  Bose condensates}},\ }\href {https://doi.org/10.1103/PhysRevB.103.035145}
  {\bibfield  {journal} {\bibinfo  {journal} {Phys. Rev. B}\ }\textbf {\bibinfo
  {volume} {103}},\ \bibinfo {pages} {035145} (\bibinfo {year}
  {2021})}\BibitemShut {NoStop}%
\bibitem [{\citenamefont {Vi\~nas Bostr\"om}\ \emph {et~al.}(2023)\citenamefont
  {Vi\~nas Bostr\"om}, \citenamefont {Parvini}, \citenamefont {McIver},
  \citenamefont {Rubio}, \citenamefont {Kusminskiy},\ and\ \citenamefont
  {Sentef}}]{mcd}%
  \BibitemOpen
  \bibfield  {author} {\bibinfo {author} {\bibfnamefont {E.}~\bibnamefont
  {Vi\~nas Bostr\"om}}, \bibinfo {author} {\bibfnamefont {T.~S.}\ \bibnamefont
  {Parvini}}, \bibinfo {author} {\bibfnamefont {J.~W.}\ \bibnamefont {McIver}},
  \bibinfo {author} {\bibfnamefont {A.}~\bibnamefont {Rubio}}, \bibinfo
  {author} {\bibfnamefont {S.~V.}\ \bibnamefont {Kusminskiy}},\ and\ \bibinfo
  {author} {\bibfnamefont {M.~A.}\ \bibnamefont {Sentef}},\ }\bibfield  {title}
  {\bibinfo {title} {Direct optical probe of magnon topology in two-dimensional
  quantum magnets},\ }\href {https://doi.org/10.1103/PhysRevLett.130.026701}
  {\bibfield  {journal} {\bibinfo  {journal} {Phys. Rev. Lett.}\ }\textbf
  {\bibinfo {volume} {130}},\ \bibinfo {pages} {026701} (\bibinfo {year}
  {2023})}\BibitemShut {NoStop}%
\bibitem [{\citenamefont {Liang}\ \emph {et~al.}(2024)\citenamefont {Liang},
  \citenamefont {Liu}, \citenamefont {Yang}, \citenamefont {Huang},
  \citenamefont {Wurstbauer}, \citenamefont {Dean}, \citenamefont {West},
  \citenamefont {Pfeiffer}, \citenamefont {Du},\ and\ \citenamefont
  {Pinczuk}}]{gravitons}%
  \BibitemOpen
  \bibfield  {author} {\bibinfo {author} {\bibfnamefont {J.}~\bibnamefont
  {Liang}}, \bibinfo {author} {\bibfnamefont {Z.}~\bibnamefont {Liu}}, \bibinfo
  {author} {\bibfnamefont {Z.}~\bibnamefont {Yang}}, \bibinfo {author}
  {\bibfnamefont {Y.}~\bibnamefont {Huang}}, \bibinfo {author} {\bibfnamefont
  {U.}~\bibnamefont {Wurstbauer}}, \bibinfo {author} {\bibfnamefont {C.~R.}\
  \bibnamefont {Dean}}, \bibinfo {author} {\bibfnamefont {K.~W.}\ \bibnamefont
  {West}}, \bibinfo {author} {\bibfnamefont {L.~N.}\ \bibnamefont {Pfeiffer}},
  \bibinfo {author} {\bibfnamefont {L.}~\bibnamefont {Du}},\ and\ \bibinfo
  {author} {\bibfnamefont {A.}~\bibnamefont {Pinczuk}},\ }\bibfield  {title}
  {\bibinfo {title} {Evidence for chiral graviton modes in fractional quantum
  hall liquids},\ }\href@noop {} {\bibfield  {journal} {\bibinfo  {journal}
  {Nature}\ }\textbf {\bibinfo {volume} {628}},\ \bibinfo {pages} {78}
  (\bibinfo {year} {2024})}\BibitemShut {NoStop}%
\bibitem [{\citenamefont {Sears}\ \emph {et~al.}(2008)\citenamefont {Sears},
  \citenamefont {Chern}, \citenamefont {Kim}, \citenamefont {Bereciartua},
  \citenamefont {Francoual}, \citenamefont {Kim},\ and\ \citenamefont
  {Kim}}]{ybkim}%
  \BibitemOpen
  \bibfield  {author} {\bibinfo {author} {\bibfnamefont {J.~A.}\ \bibnamefont
  {Sears}}, \bibinfo {author} {\bibfnamefont {L.~E.}\ \bibnamefont {Chern}},
  \bibinfo {author} {\bibfnamefont {S.}~\bibnamefont {Kim}}, \bibinfo {author}
  {\bibfnamefont {P.~J.}\ \bibnamefont {Bereciartua}}, \bibinfo {author}
  {\bibfnamefont {S.}~\bibnamefont {Francoual}}, \bibinfo {author}
  {\bibfnamefont {Y.~B.}\ \bibnamefont {Kim}},\ and\ \bibinfo {author}
  {\bibfnamefont {Y.-J.}\ \bibnamefont {Kim}},\ }\bibfield  {title} {\bibinfo
  {title} {{Ferromagnetic Kitaev interaction and the origin of large magnetic
  anisotropy in alpha-$RuCl_3$}},\ }\href@noop {} {\bibfield  {journal}
  {\bibinfo  {journal} {Nature Physics}\ }\textbf {\bibinfo {volume} {16}}
  (\bibinfo {year} {2020/08//})}\BibitemShut {NoStop}%
\bibitem [{\citenamefont {Gu}\ \emph {et~al.}(2024)\citenamefont {Gu},
  \citenamefont {Gu}, \citenamefont {Liu}, \citenamefont {Ohira-Kawamura},
  \citenamefont {Murai},\ and\ \citenamefont {Zhao}}]{PhysRevLett.132.246702}%
  \BibitemOpen
  \bibfield  {author} {\bibinfo {author} {\bibfnamefont {Y.}~\bibnamefont
  {Gu}}, \bibinfo {author} {\bibfnamefont {Y.}~\bibnamefont {Gu}}, \bibinfo
  {author} {\bibfnamefont {F.}~\bibnamefont {Liu}}, \bibinfo {author}
  {\bibfnamefont {S.}~\bibnamefont {Ohira-Kawamura}}, \bibinfo {author}
  {\bibfnamefont {N.}~\bibnamefont {Murai}},\ and\ \bibinfo {author}
  {\bibfnamefont {J.}~\bibnamefont {Zhao}},\ }\bibfield  {title} {\bibinfo
  {title} {Signatures of kitaev interactions in the van der waals ferromagnet
  ${\mathrm{vi}}_{3}$},\ }\href
  {https://doi.org/10.1103/PhysRevLett.132.246702} {\bibfield  {journal}
  {\bibinfo  {journal} {Phys. Rev. Lett.}\ }\textbf {\bibinfo {volume} {132}},\
  \bibinfo {pages} {246702} (\bibinfo {year} {2024})}\BibitemShut {NoStop}%
\bibitem [{\citenamefont {Chen}\ and\ \citenamefont
  {Villadiego}(2024)}]{chen2024}%
  \BibitemOpen
  \bibfield  {author} {\bibinfo {author} {\bibfnamefont {C.}~\bibnamefont
  {Chen}}\ and\ \bibinfo {author} {\bibfnamefont {I.~S.}\ \bibnamefont
  {Villadiego}},\ }\href {https://arxiv.org/abs/2411.08105} {\bibinfo {title}
  {Anyon polarons as a window into the competing phases of the kitaev honeycomb
  model under a zeeman field}} (\bibinfo {year} {2024}),\ \Eprint
  {https://arxiv.org/abs/2411.08105} {arXiv:2411.08105 [cond-mat.str-el]}
  \BibitemShut {NoStop}%
\bibitem [{\citenamefont {Koller}\ \emph {et~al.}(2025)\citenamefont {Koller},
  \citenamefont {Leeb}, \citenamefont {Perkins},\ and\ \citenamefont
  {Knolle}}]{koller2025}%
  \BibitemOpen
  \bibfield  {author} {\bibinfo {author} {\bibfnamefont {E.}~\bibnamefont
  {Koller}}, \bibinfo {author} {\bibfnamefont {V.}~\bibnamefont {Leeb}},
  \bibinfo {author} {\bibfnamefont {N.~B.}\ \bibnamefont {Perkins}},\ and\
  \bibinfo {author} {\bibfnamefont {J.}~\bibnamefont {Knolle}},\ }\href
  {https://arxiv.org/abs/2503.14091} {\bibinfo {title} {Raman circular
  dichroism and quantum geometry of chiral quantum spin liquids}} (\bibinfo
  {year} {2025}),\ \Eprint {https://arxiv.org/abs/2503.14091} {arXiv:2503.14091
  [cond-mat.str-el]} \BibitemShut {NoStop}%
\bibitem [{\citenamefont {Hickey}\ and\ \citenamefont {Trebst}()}]{ciaran}%
  \BibitemOpen
  \bibfield  {author} {\bibinfo {author} {\bibfnamefont {C.}~\bibnamefont
  {Hickey}}\ and\ \bibinfo {author} {\bibfnamefont {S.}~\bibnamefont
  {Trebst}},\ }\bibfield  {title} {\bibinfo {title} {{Emergence of a
  field-driven U (1) spin liquid in the Kitaev honeycomb model}},\ }\href@noop
  {} {\bibfield  {journal} {\bibinfo  {journal} {Nature Communications}\
  }\textbf {\bibinfo {volume} {10}}}\BibitemShut {NoStop}%
\bibitem [{\citenamefont {Ring}\ and\ \citenamefont
  {Schuck}(2004)}]{ring2004nuclear}%
  \BibitemOpen
  \bibfield  {author} {\bibinfo {author} {\bibfnamefont {P.}~\bibnamefont
  {Ring}}\ and\ \bibinfo {author} {\bibfnamefont {P.}~\bibnamefont {Schuck}},\
  }\href@noop {} {\emph {\bibinfo {title} {The nuclear many-body problem}}}\
  (\bibinfo  {publisher} {Springer Science \& Business Media},\ \bibinfo {year}
  {2004})\BibitemShut {NoStop}%
\bibitem [{\citenamefont {Mizusaki}\ and\ \citenamefont
  {Oi}(2012)}]{MIZUSAKI2012219}%
  \BibitemOpen
  \bibfield  {author} {\bibinfo {author} {\bibfnamefont {T.}~\bibnamefont
  {Mizusaki}}\ and\ \bibinfo {author} {\bibfnamefont {M.}~\bibnamefont {Oi}},\
  }\bibfield  {title} {\bibinfo {title} {A new formulation to calculate general
  hfb matrix elements through the pfaffian},\ }\href
  {https://doi.org/https://doi.org/10.1016/j.physletb.2012.07.023} {\bibfield
  {journal} {\bibinfo  {journal} {Physics Letters B}\ }\textbf {\bibinfo
  {volume} {715}},\ \bibinfo {pages} {219} (\bibinfo {year}
  {2012})}\BibitemShut {NoStop}%
\end{thebibliography}%

\end{document}